\documentclass[preprint]{aastex}
\begin{document}

 \title{Orbital masses of nearby luminous galaxies}
\author{Igor D. Karachentsev}
\affil{Special  Astrophysical  Observatory,  Russian  Academy  of  Sciences, Russia}

\author{Yuri N.Kudrya}
\affil{Taras Shevchenko National University of Kyiv, Ukraine}

\email{ikar@sao.ru}

\begin{abstract}

   We use observational properties of galaxies accumulated in the
Updated Nearby Galaxy Catalog to derive a dark matter mass of
 luminous galaxies via motions of their companions. The data on
orbital-to-stellar mass ratio are presented for 15 luminous galaxies
situated within 11 Mpc from us: the Milky Way, M31, M81, NGC5128,
IC342, NGC253, NGC4736, NGC5236, NGC6946, M101, NGC4258, NGC4594,
NGC3115, NGC3627 and NGC3368, as well as for a composit suite around
other nearby galaxies of moderate and low luminosity. The typical
ratio for them is $M_{orb}/M_*$ = 31, corresponding to the mean local
density of matter $\Omega_m$ = 0.09, i.e  1/3 of the global cosmic
density. This quantity seems to be rather an upper limit of dark
matter density, since the peripheric population of the suites may suffer
from the presence of fictitious unbound members. We notice that the
Milky Way and M31 haloes have lower dimensions and lower stellar masses than
those of other 13 nearby  luminous galaxies. However, the dark-to-stellar
mass ratio for both the Milky Way and M31 is the typical one for other 
neighboring luminous galaxies. The distortion in the Hubble flow,
observed around the Local Group and five other neighboring groups yields
their total masses within the radius of zero velocity surface,$R_0$, which are
slightly lower than the orbital and virial values. This difference may be due
to the effect of dark energy, producing a kind of ``mass defect'' within $R_0$.


 \end{abstract}


 \section{Introduction}
In spite of  tremendous success of observational cosmology, reached
over the past quarter century,
many issues regarding the nature of dark matter and its
distribution in the universe relative to the visible (stellar) matter
still remain unresolved.
Numerous studies (Karachentsev 1966, Rood et al. 1970, Bahcall et al. 2000)
have shown that in  groups and clusters of galaxies the ratio of
dark (virial) mass  to  stellar mass systematically increases
with  size and population of a  given system of galaxies.  In the richest
clusters, such as the Coma, the
$M_{DM}/M_*$ ratio reaches up to two orders of magnitude. If all the
galaxies are part of clusters,
dark matter associated with them would provide the average  density of
matter in space  amounting to $\Omega_m\simeq0.26$
(Bahcall \& Kulier, 2014), corresponding to the standard cosmological
$\Lambda$CDM  model  (Spergel et al. 2007).

However, no more than 10\% of all galaxies belong to  rich clusters
(Libeskind et al, 2013, Cautun et al. 2014). Most of them are included
in groups of different multiplicity, which are concentrated in the filaments
and ``sheets'', forming a large-scale ``cosmic web''
( Bond et al. 1996, Shandarin et al. 2004, Einasto et al. 2011).
Looking at the data on 11000 galaxies of the nearby universe
with radial velocities $V_{LG}\leq 3500$~km$\:$s$^{-1}$ , Makarov \&
Karachentsev (2011) have identified in this
volume about 400 groups and clusters of galaxies and determined their
virial masses. The summation of  virial
masses of groups and clusters in the volume of $\sim$50~Mpc radius led
to the average density estimate of
$\Omega_m$(local)$\simeq 0.08\pm0.02$,
which is three times lower than the global cosmic density. This result
confirmed the earlier estimates of $\Omega_m\sim(0.08-0.10$), which were
obtained for the Local universe by
Vennik (1984), Tully (1987), Magtesian (1988)  and other authors.  A
threefold difference between the estimates of  $\Omega_m$(local) and
$\Omega_m$(global) did  not cause much concern among theorists.
It was considered quite obvious that  dark matter is not distributed in
clusters and
groups with the same concentration  as  stellar matter (biasing effect).
Darker peripheries
of the clusters probably contain a large amount of dark matter, the
presence of which eliminates the paradox of
``missing dark matter''.

The assumption of massive dark halos existing around the clusters and groups
of galaxies is not, however, confirmed by the observations. Investigating
the Hubble flow of galaxies around the Virgo, the nearest cluster of galaxies,
Karachentsev et al. (2014b) showed that the total mass of the cluster,
determined from the external motions  of galaxies is in a good agreement
with the virial mass estimate based on the  motions within the cluster.
Since the total mass of the  Virgo cluster was estimated on a scale of the
``zero velocity sphere'' radius, $R_0$, which is $\sim$3.7 times larger than
the virial radius $R_v$,  this result
gives evidence against the localization of a significant amount of dark
matter in the layer between $R_v$ and $R_0$.
A similar situation occurs  around the Local Group of galaxies (Karachentsev
et al. 2009). Consequently, we should be on the outlook for other ideas and
observational data  to resolve the paradox of missing dark matter.

The recently published ``Updated Nearby Galaxy Catalog'' = UNGC, (Karachentsev
et al., 2013) contains a summary of data on
radial velocities, distances and other observable parameters of about
800 galaxies located within a
11~Mpc radius around us. More than 300 galaxies of this sample have
accurate distance measurements  with a better than 10\% accuracy  obtained by 
the Tip of the Red Giant Branch  from observations with
the Hubble Space Telescope. Due to the proximity of the
UNGC- objects,  the kinematic data density in the catalog proves to be 6
times higher than
in the sample of the nearby ($D\leq50$ Mpc) universe (Makarov \& Karachentsev,
2011, hereafter MK11). This circumstance,
and the presence of  individual distance measurements in many  UNGC galaxies
allows us to investigate the structure
of nearby groups and their vicinities with unprecedented detail. Determining
the masses of the most nearby galaxies
from the motions of their companions  is the main subject of this paper.

\section{Projected and orbital mass estimates}

To determine the mass of a system of N point-like bodies, one usually uses  the
virial theorem in the form of
$$M_v=(3\pi/2)\times G^{-1}\times S^2_v\times R^{-1}_h, \eqno(1) $$
where $G$ is the gravitational constant, $S^2_v$ is the  velocity dispersion
on the line of sight, $R_h$ is the average
harmonic separation between the group members in the projection on the sky,
and $(3\pi/2)$ is the average
projection factor at  arbitrary group orientation with respect to the
line of sight  (Bahcall \& Tremaine, 1981). But this estimator is
statistically offset and inefficient. Therefore, Heisler et al. (1985)
proposed to estimate the mass of a group in a more robust way:

$$M_p=(32/\pi)\times N \times(N-3/2)^{-1} G^{-1}\langle\Delta V^2\times
R_p\rangle, \eqno(2)$$
where $M_p$ is the so-called ``projected'' mass, N is the number of
objects, and   $\langle\Delta V^2
\times R_p\rangle$ is the average
product of squared radial velocity of the component relative to  the
group center, and its projection separation from the center.
Both these mass estimators presume spherical symmetry of the groups
as well as isotropic velocity distribution. But as shown by Wojtak (2013)
many groups are highly aspherical, with shapes approximately by nearly
prolate ellipsoids. According to Wojtak (2013) their mean spatial axial
ratio is $\sim 0.66$ and the mean axial ratio of the velocity ellipsoids
is $\sim 0.78$. Furthermore, simulated dark matter haloes tend to be
aligned with the cosmic web in the way that the semi-major axis is
aligned with the local filaments and the semi-minor axis is pointing to
neighbouring voids (Libeskind et al. 2013). Being mostly located in the
Local Sheet, the nearby groups may be preferentially observed along their
major or median axis that would have any effect on the mass estimates.

If the group is dominated by a massive galaxy, surrounded by a set
of test particles with random orientation of their orbits, one can use
the  ``orbital'' mass estimate  (Karachentsev, 2005):

$$M_{orb} = (32/3\pi)(1-2e^2/3)^{-1}\times  G^{-1}\times \langle\Delta
V^2_{12}\times R_{p 12}\rangle, \eqno(3)$$

where $\Delta V_{12}$ and  $R_{p 12}$ are the velocity difference and
the projected separation of companions relative to the main
galaxy, and $ e $ is the prevailing orbit eccentricity. Assuming the
typical  eccentricity value of
$\langle e^2\rangle\simeq1/2$ (Barber et al. 2014), we get

$$M_{orb} = (16/\pi)\times G^{-1} \times \langle\Delta V^2_{12}\times
R_{p 12}\rangle. \eqno(4)$$

 For completeness, we also mention another approach to  mass estimation
proposed by Beloborodov \& Levin (2004). Based on the natural assumption
that companions of the main galaxy are observed at random orbital phase moments,
they offered so-called ``orbital roulette estimator''

$$M_{rlt}= 6\times(2-\langle e^2\rangle)^{-1}\times G^{-1}\times
\langle\Delta V^2_{12}\times R_{p 12}\rangle, \eqno(5)$$

 which uses just the same obervables, but yields at 
$\langle e^2\rangle=1/2$ the mass estimate 21\% smaller than (4).
Note that at $N=2$ the projected mass estimate (2) coincides with the
orbital estimate (4). We will use the orbital mass estimator further on.

\section{Neighboring giants and their suites}

Possessing the data on the distances and luminosities of 869 galaxies of
the Local Volume, Karachentsev et al. (2013) have determined for each
galaxy its tidal index

$$\Theta_1 = \max[\log(M_n^*/D_n^3)]+C, \,\,\, n=1, 2,... N, \eqno(6) $$
where $M^*$ is the stellar mass of the neighboring galaxy, and $D_n$ is
its spatial separation from the considered galaxy.
The stellar mass of the galaxy was assumed to be equal to its K-band
luminosity at  $M^*/L_K=1 M_{\odot}/L_{\odot}$ (Bell et al, 2003).
Ranking the surrounding galaxies by the magnitude of their tidal force,
$F_n\sim M^*/D_n^3$, allowed  to find the most influential neighbor,
called the Main Disturber (= MD). Here the ratio of the total mass of
the galaxy to its stellar mass was considered to be constant regardless
of the
luminosity and morphology of galaxies.  The constant C = -10.96 in equation
(6) was chosen so that the galaxy with $\Theta_1=0$ was located at the
``zero  velocity  sphere''
relative to its MD. In other words, the galaxy with $\Theta_1>0$ was
regarded as causally (gravitationally) related to its MD as
their crossing time was shorter than the age of the universe,
$T_0=13.7$ Gyr.   Consequently, the causally unrelated galaxies with
$\Theta_1<0$  were referred to as  the  population of ``general
field''.

Obviously, the galaxies  which have a common MD can be combined in a
certain association, or a MD ``suite''.
At that, an aggregate of suite members  with positive  $\Theta_1$ values
is quite consistent with the notion of a physically bound group of
galaxies. Karachentsev et al. (2014a) have analyzed  different properties
of galaxies in the suites, as well as properties
of their main galaxies. As expected, the most massive MDs possess the
most populous suites. The total number of companions around 15 most massive
galaxies makes up about a half of the total population of the Local Volume.

The full list of suites, ranked by the number of suite members from n = 53
to n = 1 is presented in Table 1 (Karachentsev et al. 2014a). \footnote{Its
machine--readable version is available at
http://lv.sao.ru/lvgdb/article/suites\_dw\_Table1.txt.} The Table 5 (Appendix)  below
presents the summary of 15 richest groups (suites) of the Local Volume,
in which at least 6 galaxies  have measured radial velocities.
We did not include in the  Table 5  those members of suites, which radial
velocities remain still unmeasured. These cases consist of about 1/3 of
the total amount of suite members.

The heading line of each suite presents: the name of the Main Disturber,
its distance in Mpc, its stellar mass and the  value of  orbital
mass with the standard error. The suites (groups) are arranged
in the descending order of their total population. The following is given for the
members of each suite: (1) name of the galaxy in UNGC catalog;
(2) the tidal index $\Theta_1$ by which the members of the suite are ranked;
(3) projection separation of the suite member  from the MD  in kpc,
assuming that all the companions of the MD
are at the same distance from the observer as the  MD itself;
(4) absolute value of the  radial velocity difference  of member
of the suite and the  MD in km/s.

The distribution of 351 companions by the  radial velocity  difference and
projection separation relative to their main galaxies is presented in
three panels of Fig.1. The upper panel of the figure shows the
\{$|\Delta V|, R_p$\}  diagram for 31 companions of the Milky Way = MW (squares)
and 39 members of the M~31 = Andromeda suite (diamonds). The companions
of massive galaxies with the tidal index   $\Theta\geq0$,  considered
to be physical, are represented by closed symbols, while the members
of the suites with  $-0.5<\Theta_1<0$ are shown by the open symbols. The
extension of the companion sample by the objects with slightly negative
values of  $\Theta_1$  was done  not to miss some possible
physical members of the group, in which the distances are as yet measured
with low accuracy. The objects in this boundary category may appear to
be both the real companions of main galaxies or belong to the
population of general field.
Note that for the MW companions we are not listing the spatial
distances, but their projection on the plane perpendicular to the line of
sight towards the MW center.

The middle panel of Fig.1 shows the  \{$|\Delta V|, R_p$\} distribution
for 174 members of rich suites around 13 other massive nearby
galaxies. Prospective physical companions with $\Theta_1\geq0$ (N=142)
are also marked here by solid symbols.

In addition to 15 rich suites,  the Local Volume comprises a lot of small
suites, where the radial velocities are measured in one or several
presumed companions. We have combined these small suites  in a composite
(``synthetic'') suite.
The   \{$|\Delta V|, R_p$\} diagram
for 107 companions  uniting small suites is represented on the lower
panel of Fig. 1. At that, we only kept the  cases where the stellar mass
of companion does not exceed half the mass of the main galaxy.

The dashed lines in all the three panels of Fig.1 show  quadratic
regressions of the velocity difference on the projection separation of
companions. For the suites of galaxies around MW and M31, the  regression
has a negative slope. While for the synthetic suite of 142 companions around the
13 most massive galaxies and
for the synthetic suite, uniting small suites,  regressions show a weak increase
in velocity dispersion from the center to the suite periphery.
Different behavior of the regressions  may indicate the atypical character of
motion of  the MW and M31 companions in comparison to the suites of other
massive nearby galaxies. Another reason of the rising part of the
velocity dispersion may be caused by the presence of large scale halo when
nearby groups are a part of larger structure, the Local Sheet, which would
give rise to the observed enhancement of the velocity dispersion at large
radii. However, a more obvious reason for this phenomenon
is caused by the presence  on the suite outskirts of an admixture of
some false members entering the suites from the general field.

The basic characteristics of the considered suites are presented in Table 1.
Its columns contain: (1) name of the suite/group by its main galaxy,
(2) the number of physical ($\Theta_1\geq0$) members of the group with
measured radial velocities,
(3)  the average projection separation of the companions from the main galaxy
(kpc),
(4)  the mean absolute value of the  radial velocity difference of the
companions relative to the main galaxy
(km/s),
(5) the main galaxy stellar mass in the units of  $10^{10}M_{\odot}$,
(6) the  value of orbital mass of the group
(suite) in the units of $10^{12}M_{\odot}$ and its standard error
coming from the error of the  mean in equation (4). 
The location of suites in Table 1 corresponds to their breakdown in the
three panels of Fig. 1:
the first lines contain  the data for the MW and M31 groups,
followed by  the characteristics of 13 other most populated groups of the
Local Volume, and   the end of the table shows the average parameters of 
composite suite. Since the main galaxies in the composite suite significantly
differ by their stellar mass, we have divided the synthetic suite  into
three subsamples having about the same number of companions with measured
radial velocities.

 Distribution of the surface number density of 297 companions along the 
radius of the combined suite is presented in Fig. 2 in the log-log scale. 
The solid circles correspond to the number of physical companions, and the
crosses consider the additional number of suite members with   
$\Theta_1=[0, -0.5]$. The vertical bars show the statistical error of  
$\sqrt{N-1}$. The dashed line represents the quadratic regression

  $$\log\Sigma(R_p) = -3.88 - 2.18 \ x - 0.56 \ x^2 ,$$
where $x = \log (R_p/200$ kpc).

As can be seen, the surface number density profile for the synthetic suite 
is well compatible with the radial profile of the surface mass density for
the standard NFW-profile of the dark halo (Navarro et al. 1997, Wang et al. 2014), 
as given by equation (41) and Figure 8 in Lokas \& Mamon (2001).

     \section{Milky Way and Andromeda suites as compared with others}

Modeling the structure and kinematics of galaxy groups within the
$\Lambda$CDM  paradigm, many authors (Libeskind et al. 2010,
Zavala et al. 2009, Knebe et al. 2011) choose the Local Group
to make a comparison with the observational data.
As known, the Local Group has two gravitating centers:
the MW and M31,
which are approaching each other with mutual velocity of about
110 km/s. This binary character is not an exclusive feature. For example,
the neighboring
groups: M81 and NGC~2403, IC~342 and Maffei~I, NGC~5128 and NGC~5236 also
belong to the class of binary merging groups.  But from the standpoint of
the group mass estimate
from the orbital motions  of the companions,  the listed galaxies have to
be considered as standalone dynamical centers.

Previously Karachentsev et al. (2014a) noted that judging on some
morphological features  the groups of galaxies around the MW
and M31 are not quite typical. This primarily refers
to the presence near the MW of two companions (Magellanic Clouds)
rich in gas.  There are also other features that distinguish the  MW and
M31 groups among other nearby ones.

Six histograms of Fig.3
represent the distributions of 15 most populated suites in the Local
Volume based on the following parameters:
the average projected separation of the companions from the main galaxy,
$\langle R_p\rangle$, the mean absolute value of the  radial velocity
difference of the companion and the main galaxy,
the logarithm of stellar mass of the MD,
the MD orbital mass,
the ratio of the orbital mass-to-sum of stellar masses of all the
galaxies in the group, $M_{orb}/\Sigma M_*$,
and the average crossing time  $t_{cr}=R_p/\sigma_v$  for the suite members
, where $t_{cr}$ is expressed in terms of the age of the universe, $T_0=13.7$  Gyr.
The groups of galaxies around the MW and Andromeda are marked with
``M'' and ``A'', respectively.

According to these data, the linear dimension of the suites around MW and
M31 are approximately 2 times less extended than the typical suite of other
neighboring massive galaxies. In the case of MW, that can be caused by
the obvious selection effect: most of the recently discovered ultra-low
luminosity companions of the MW  were found at the distances of  less than
100 kpc (Willman et al. 2005, Belokurov et al. 2006). To some extent, the small
linear size of the suite of companions around M31 can also be caused by a
selection effect, since the most thorough search for new companions was
carried out in a limited region around M31  (Ibata et al. 2007, Martin
et al. 2009). However, the most plausible explanation of this difference
may also be the presence in the  suites of neighboring massive galaxies
of a certain number of false members, which appear on the periphery of
the suites from the general field.

In contrast to the linear dimensions, radial velocity dispersion for the
companions of MW and M31 does not stand out among the other groups
(panel ``b'').

The ``c''  histogram data  show that based on their stellar masses,
both
MW and M31 do not get in the top ten most
massive galaxies of the Local Volume.  This may be also the
reason of understated linear dimensions of the suites around the MW and M31.

The ``d''  and  ``e''  histograms show the distribution of 15 suites
by the orbital mass  and by the  ratio of the orbital
mass-to-sum of stellar  masses of the  group members, respectively.
The two   groups located most rightward on these panels correspond to
the suite around NGC~4594 (``Sombrero'')
and the group NGC~3368/3379 (Leo~I).

If in the distribution of  suites by the value of $M_{orb}$ both groups
MW and M31 are shifted towards
the lower values relative to the average, whereas based on the
$M_{orb}/\Sigma M_*$  parameter, both groups are not significantly different
from the rest.

The lower panel of Fig.3 shows the distribution of suites by the
average crossing time of the companions.
A typical dynamic
situation in the group of the Local Volume is expressed by
the fact that the companions of massive galaxies have time to
make about 5 oscillations around the center, which is sufficient for the
group to get virialized.
Two suites on the right side of the histogram with  $t_{cr} \sim 1/2$ are
the scattered groups around NGC~253 (the Sculptor filament) and NGC~4736
(the CVn~I cloud), the dynamical relaxation of which has apparently not yet
achieved.

\section{Orbital and projected masses of neighboring groups}

As noted above, the formation of suites around the nearby galaxies was
made based on the data on mutual separations and stellar masses
($L_K$-luminosities)  of galaxies in the Local Volume.   Radial
velocities of galaxies were not taken into account here. Among $\sim400$
groups from   the list of MK11 there are fairly nearby groups falling into
the Local Volume. In 18 of them the number of members with known radial
velocities
is not too small  $(N_v\geq4)$ to estimate the projected mass
of the group with an acceptable statistical error.
The sample of these 18 groups presents a unique opportunity to compare
the dynamical mass estimates made applying different methods to the systems
of galaxies, the  principles of identification of which were essentially
different.

Let us recall that the arrangement of galaxies in  MK-groups was carried
out via the  pairwise revision of all galaxies with two conditions:
the total energy of a virtual bound pair must be negative, and the pair
components have to be within the  ``zero velocity sphere,''
determined by the total mass of the pair. In the space of  projected
separations
$R_p$ and radial velocity differences $\Delta V_{12}$, these conditions
are expressed as

$$\Delta V^2_{12} R_p< 2 G (M_1+M_2), \eqno(7)$$
$$\pi H_0^2 R_p< 8 G(M_1+M_2), \eqno(8)$$
where  the condition

$$M/M_*=\kappa=6 \eqno(9)$$

was assumed for the relation of the dynamical mass of each galaxy to
its stellar mass.

Then, all  the virtual bound pairs with common members were united
in a group. Unlike
another widely used method of organizing  the galaxies in
``friends of friends'' groups (Huchra \& Geller 1982, Crook et al. 2007),
the (7) -- (9) criterion contains only one arbitrary
dimensionless parameter  $\kappa$.
At the empirically selected value of $\kappa=6$,
the  (7)--(9) criterion brings together in pairs, groups and
clusters about 54\% of all galaxies, what is in good agreement with the
observed structure of the Local Volume (see the details in the MK11).

A comparison of parameters of the suites around 18  nearby
massive   galaxies with the characteristics of the corresponding nearby
MK-groups is given in Table 2. The top rows of the table represent the
data for the MK-groups, while  the lower rows list the  parameters of the
suites. The columns contain:
(1)  name of the main galaxy of the group/suite;
(2)  number of galaxies in the group/suite with measured radial
velocities;
(3)  the distance to the group (Mpc), determined by the mean radial
velocity  of the group members relative to the Local Group centroid
at  $H_0=73$ km/s/Mpc, and the individual distance of the
principal galaxy of the suite;
(4)  dispersion of radial velocities in the group and the mean-square
difference of the companion velocities  relative to
the main galaxy (km/s);
(5)  the mean harmonic  radius of the group and the
mean projection separation of companions from the main galaxy
(kpc);
(6)  logarithm of the total stellar mass of the group or the suite (in
$M_{\odot}$);
(7) logarithm of the projected mass of the group and the orbital mass
of the suite (in   $M_{\odot}$);
(8) the ratio of the projected (or orbital) mass-to-total stellar mass
in the logarithmic scale;
(9) morphological type of the main galaxy on de Vaucouleurs scale;
(10) difference between apparent K-magnitudes of the first and
second members of the group;
(11--13)  the tidal indices, characterizing the environment density
of the main galaxy in the group: here the $\Theta_1$ index,
determined by equation (6), expresses the contribution of the most
significant  neighbor, the  $\Theta_5$ index
accounts for the effect of five important neighbors, while
the  $\Theta_J$ index corresponds to the logarithm of stellar density
contrast in a sphere of  1 Mpc radius
around the main galaxy taken with respect to the mean cosmic density.
The last line in the table shows the mean values  of the considered quantities.
Note that the luminosity of the brightest suite member does not exceed 1/4 of
the MD's luminosity for 10 of the 15 suites, that justifys the consideration of
suite galaxies as test particles orbiting around the central massive body.

One can notice that Table 2 has no data on the groups around IC~342 and NGC~6946.
They are not included in the list of MK-groups because located in the
zone of strong Galactic extinction. The groups of companions around the
MW and M31 are also missed because their distances based on
the mean radial velocities of the galaxies, used by Makarov \&Karachentsev
(2011), would have no physical meaning. A comparison of  the Table 2 data on
the groups versus the suites reveals the following properties.

a) The total number of galaxies in the MK-groups, 227, is comparable
to the total number of physical members of the suites: 170 at   $\Theta_1>0$
and 224 at $\Theta_1>-0.5$.
Consequently,  the association of galaxies into suites by the
zones of gravitational influence around  dominant galaxies, and by
the MK-criterion (7--9) have approximately the same clustering efficiency
rate. However,  the data presented  reveal also significant
individual differences in the populations of groups and suites.
For example, in the   NGC~891, NGC~4631 and NGC~4736 groups, this ratio
amounts to 18:4, 28:5 and 5:15, respectively. The greatest differences
are typical for the scattered groups (suites), where the second member of
the group by luminosity competes with the MD.

b) The mean radial velocity dispersion in groups, equal to 83 km/s,
and  the mean square velocity difference of the companions in the suites,
amounting to 99 km/s, are in a reasonable agreement with each other.
In other words,  condition (7) in the MK-criterion does not possess
strong selectivity against the pairs of galaxies with a large radial
velocity difference.

c) The difference between the Hubble distance to the  groups,
$D_H =\langle V_{LG}\rangle /H_0$,  and individually measured distance to
the main galaxy of the suite, $D_ {MD} $, is on the
average small: 7.44 Mpc and 7.39 Mpc, respectively. While in
some groups, for instance, in NGC~628 and NGC~2903, these distances differ
by half (due to the bulk motions towards the  Virgo cluster), which
affects the luminosity of the group and influences the
number of clustered members in it.

d) Individual differences between the estimates of the projected
mass and  orbital mass   are quite large. In the case of groups of galaxies
around NGC~253, NGC~891, NGC~3627, NGC~4594, NGC~4736 and NGC~5194, these
differences exceed the factor 3.  Nevertheless, the average values of
$\langle\log M_p\rangle=12.44$ and $\langle\log M_{orb}\rangle=12.41$  for
an ensemble of 18 groups/suites are in good agreement with each other.
Similarly, the average ratios of  $\langle\log(M_p/\Sigma M^*)\rangle = 1.50$
and $\langle\log(M_{orb}/\Sigma M^*)\rangle =1.53$  do not show any significant
systematic difference, although in some groups/suites these ratios
differ significantly.
In addition to random factors caused by the poor statistics, the differences
in the estimates of $M_ {orb}$ and $M_p$ occur more in  scattered groups, where we
can discern the substructures around the galaxies, which are only slightly
less massive than the main member of the group. The examples revealing the
presence of such hierarchical substructures can be found in  the
NGC~891/NGC~925, NGC~3368/NGC~3379 and NGC~5194/NGC~5055 groups.

e) The data of the last columns of Table 2 show that the density of the
group environment, the difference in the  apparent magnitudes of  two
brightest members of the group, and the morphological type of the main
galaxy do not affect  the ratio of dark-to-luminous matter in the group
in  a substantial way.

 It should be emphasized that the derived above agreement between
the typical values of orbital and projected masses in the Local Volume,
$M_{orb}\simeq M_p \simeq 33 M^*$, is not trivial one. The UNGC catalog
contains approximately the same number of radial velocities as what was 
used by MK11 within $D < 11$ Mpc. However, the UNGC has much more data on
galaxy distances than MK11 sample. Actually, MK11 estimated distances
to a group via the average redshift of its members burdened by peculiar
velocities and local streams. This is does not matter in the case of
UNGC which collected hundreds accurate individual distances. Another
significant difference is caused by different algorithms applied to the
galaxy grouping. To find a group, MK11 used as separations and luminosities
of galaxies, as well their radial velocities. In the case of UNGC, only
3D-separations and luminosities (but not redshifts) were used to identify
a suite of companions around a dominant galaxy. We can not state that one 
finding algorithm is better (or objective) than the other. But they both
yield almost the same average ratio $ M_{DM}/M^*$ for the small local 
structures.

\section{Orbital-to-stellar mass ratios}

The orbital mass estimates for the populated suites, shown in Tables 1 and
2, were determined by the suite members  with tidal indices $\Theta_1\geq0$.
Obviously, the choice of the maximum value of $\Theta_1$, based on which
the galaxies were included in the suite,
affects the number of members of the suite, their total luminosity,
and the orbital mass estimate. With large positive values of $\Theta_1$,
many physical  companions of the MD do not make it  in its suite. The
orbital mass would in this case prove to be
underestimated. On the contrary,  inclusion of galaxies with arbitrary
negative values of $\Theta_1$ in the suite
contributes to its pollution with false members from the general
field, and thus leads to an overestimation of the $M_{orb}$.

Figure 4 shows how sensitive are the $M_{orb}/\Sigma M_*$  estimates
to the choice of the threshold value of $\Theta_1$  for 15  most populous
groups in the Local Volume. The variations of the $M_{orb}/\Sigma M_*$
ratio depending on $\Theta_1$ in the range of   $[-0.5<\Theta_1<0.5]$ with
the increments of 0.1 are shown in this figure for each of the 15 suites.

We can see from these tracks that a rapid growth of $\log(M_{orb}/\Sigma M_*$)
towards negative values  of $\Theta_1$ takes place in 5 suites only:
NGC~253, NGC~4258, NGC~4594, NGC~4736, and NGC~5236. In the remaining
groups (suites), the ratio of the orbital mass-to-sum of stellar masses
is weakly responsive to the variation of $\Theta_1$. This is true in
particular for the companions of the suites around the MW and M31,
which are marked in the  figure by solid diamonds and squares, respectively.
The logarithm of the average value of $\langle M_{orb}/\Sigma M_*\rangle$
for all the groups,
shown by open diamonds, varies within the range of [1.59 -- 1.67] when the
threshold of $\Theta_1$ changes from +0.4 to --0.4.

According to Jones et al. (2006), the mean density of  stellar mass
amounts to $j_*=4.28\times 10^8 (M_{\odot}$/Mpc$^3$) at $H_0=73$ km/s/Mpc.
 Assuming  $M_*/L_K = 1 M_{\odot}/L_{\odot}$  (Bell et al, 2003),
the mean cosmic density of matter $\Omega_m=0.28$ in the
standard  $\Lambda$-CDM  model is expressed as $M_{DM}/M_*=97$. This value
is shown in Fig.~4 by the dashed horizontal line. As one can see, all the
groups (suites), except for the NGC~3368 (Leo~I group) and NGC~4594
(``Sombrero'') have the $M_{orb}/\Sigma M_*$ ratios below this value.
Consequently, the amount of dark matter in the suite volumes
around the massive nearby galaxies is clearly
not enough to provide the cosmic density $\Omega_m=0.28$.

The upper panel of Fig.~5 shows the distribution of 18 most populated
nearby groups given in Table 2 by their total stellar
masses and the projected mass estimates. The solid line corresponds to
the cosmic value of $M_{DM}/M_*$ = 97. As it is seen, all the
nearby groups locate
below the  $\Omega_m=0.28$ line, following the value which is
about 3 times lower (dashed line).

A similar diagram for the orbital mass estimates of the suites is shown
at the bottom panel of Figure 5. Each of the 15 populated suites (Table 1)
is shown by a circle with a vertical bar, corresponding to the standard
error of $M_ {orb}$.
The solid and dashed lines are fixing the values of $M_{DM}/M_*$ = 97
($\Omega_m=0.28$) and 31 $(\Omega_m=0.09)$, respectively. The major part
of the suites is concentrated near
the line of $\Omega_m=0.09$, and only two suites: NGC~3368 and
NGC~4594 reside above the  $\Omega_m=0.28$ line.
In addition to 15 populated suites, three small diamonds in the figure show
the average values of  $\langle M_{orb}/\Sigma M_*\rangle$  for the
synthetic suites: L, M and S, the data on which are listed at the bottom
of  Table~1.
Synthetic suites  around the galaxies of small (S) and medium (M) mass
are characterized by a high $\langle M_{orb}/\Sigma M_*\rangle$  ratio,
and within the error they lie on the line of  $\Omega_m=0.28$.

Among the smallest suites, isolated pairs of dwarf galaxies can be found,
where each component of the pair is the Main Disturber for the second
component. The list of such 12 dwarf pairs is presented in Table 3.
The low luminosity of these
galaxies clearly does not favour their detection outside the Local Volume.
Nevertheless, the catalog of binary galaxies in the Local Supercluster
by Karachentsev \&Makarov (2008) as well as the list of multiple dwarfs by
(Makarov \& Uklein 2012) contain about 50 more similar dwarf pairs.

The binary systems in Table 3 are ranked by their distance from us. The table columns 
represent  the following data, adopted from the  UNGC catalog: (1)
names of the components; (2)  the  distances (Mpc); (3)  tidal indices,
characterizing the degree of mutual gravitational influence;
(4) logarithm of stellar
mass ($M_{\odot}$); (5) logarithm of the hydrogen mass  ($M_{\odot}$);
(6)  the radial velocity difference of the components   (km/s);
(7)  the velocity measurement error  (km/s);
(8) projected separation (kpc);
(9) logarithm of orbital mass  ($M_{\odot}$).
According to these data, the average  orbital mass of dwarf pairs amounts
to  $1.83\times 10^{11} M_{\odot}$,
and the average sum of  stellar masses of the components is
$7.7\times 10^8 M_{\odot}$. The ratio of these quantities
$\langle M_{orb}\rangle /\langle M_1^*+M_2^*\rangle=237\pm172$ is  shown in
the lower panel of Fig. 5 by a square. The position of  the square
above the $M_{DM}/M_*=97$ line (but with a large error bar)
gives an impression that the  dark-to-baryonic matter ratio tends to be
higher in the low-luminosity galaxies than that in the  galaxies with normal
luminosity. This assertion has been repeatedly stated in the literature
(Mateo 1998, Moster et al. 2010).

However, we should pay attention to two important circumstances here. The
orbital mass estimates using equation (4) are statistically biased ones.
In the presence of radial velocity measurement errors, the product
$\langle\Delta V^2_{12} \times R_{p 12}\rangle$  in (4) should be replaced by
$\langle(\Delta V^2_{12} -\sigma_{v1}^2-\sigma^2_{v2}) \times R_{p 12}\rangle$.
Accounting for the  contribution of 
 $\sigma_{v1}, \sigma_{v2}$ errors lowers the average ratio of
$\langle M_{orb}\rangle/\Sigma M_*$ to $214\pm155$.

Comparison of stellar vs. hydrogen masses of dwarf galaxies in pairs
shows that these values are comparable with each other. The mean difference
$\langle\log M_{HI}-\log M_*\rangle$ from the Table 3 data is equal to --0.13.
It becomes positive, +0.14, if one takes into consideration that the mass of
gas accounting for helium and molecular hydrogen is on the average 1.85
times larger than the mass of atomic hydrogen (Fukugita \& Peebles 2004).
Having introduced both the corrections, the ratio of the orbital
mass-to-sum of baryonic masses of the pairs drops to
$\langle M_{orb}\rangle/\Sigma (M_*+ M_{gas})= 78\pm56$.

Summing the stellar masses in all the considered suites of the Local Volume,
we obtain  $\Sigma M_*= 1.52\times 10^{12} M_{\odot}$.  With the average local
stellar mass density of $j_*=6.0\times 10^8 M_{\odot}/$Mpc$^3$ (Karachentsev
et al. 2013)  and $M_*/L_K = 1 M_{\odot}/L_{\odot}$,
the sphere of  10~Mpc radius contains the total stellar
mass of  $1.88\times 10^{12}M_{\odot}$ (a small correction is introduced
here, accounting for the  zone of interstellar extinction in the Milky Way).
Therefore, the studied suites contain 80\% of the total stellar mass in the
Local Volume. The total orbital mass for them is $8.1\times 10^{13} M_{\odot}$.
The $\Sigma M_{orb}/\Sigma M_*=53$  ratio
is an important dynamic characteristic of the Local Volume.
The 20\%  of stellar mass  we have unaccounted for are distributed as
the field galaxies. They contribute
both in the denominator and the numerator of the $\Sigma M_{orb}/\Sigma M_*$
ratio, and probably have little effect on its value. The position of the
whole Local Volume in the  $\log M_{orb}\propto \log M_*$ diagram is shown by
a large open diamond in the upper right corner of the bottom panel in
Figure 5. The ratio of the sum of masses, equal to 53 is equivalent to the
average density in the Local Volume, $\Omega_m$(LV) = 0.15.

Note, however, that more than a half of the contribution to the total value
of $M_{orb}$ in the Local Volume is introduced only by two suites
around NGC~4594 and NGC~3368.
Both systems are on the far edge of the Local Volume at distances of 9.3 Mpc
and 10.4 Mpc, respectively. Individual distances to the majority of members
of both  suites
are determined with an error of   $\sim2$ Mpc  by the Tully-Fisher method,
or not measured at all, being instead   attributed the distance of the main
galaxy of the group.
Moreover, the NGC~4594 and NGC~3368 groups are located in the immediate
vicinity of the  ``zero velocity sphere''  of the  Virgo cluster, where it
is difficult to separate the bulk infall motions of galaxies toward the
cluster from virial motions within the groups.  We did not notice
anything other special about their location with respect to the Local Sheet.
Obviously, these groups need a more  comprehensive, special  analysis of their
structure and kinematics with the use of new observational data.

It should be added that out of 6 members of the   NGC~4594 suite, one galaxy,
DDO~148, resides at the  large projected distance of 1.1 Mpc from the
Sombrero galaxy, having the  radial velocity difference of 276 km/s.
The contribution of DDO~148 to the total orbital mass estimate of the Sombrero
suite is more than a half. Since the distance to
DDO~148, $D$ =9.0 Mpc is determined   by the Tully-Fisher  method with
the error of  $\pm2$ Mpc, a more accurate estimate of its distance can
dramatically change the value of
$M_{orb}$ for this suite.

If we limit the Local Volume by the 8 Mpc radius, excluding a still uncertain
situation on the far boundary, the ratio of the total orbital mass of all the
suites in this volume to the sum of stellar masses will be
$\Sigma M_{orb}/\Sigma M_*=30$  at $M_*/L_K = 1 M_{\odot}/L_{\odot}$.
On the lower panel of Fig. 5, this value,
falling on the $\Omega_m=0.09$ line is
marked by a large solid diamond.

\section{Masses derived from Hubble flow around the nearby groups}

A high density of observational data on the radial velocities and
distances of galaxies in the Local Volume gives an opportunity to determine
the masses of nearby
groups not only by the virial motions, but also by perturbations of the
Hubble flow around them. This idea was proposed by Lynden-Bell (1981) and
Sandage (1986), and is
based on the measurement of the radius of the  zero velocity sphere $R_0$
which separates a group (or a cluster) from the surrounding  volume that
expands.

In the standard
cosmological model with the parameters  $H_0=73$ km/s/Mpc and $\Omega_m=0.24$
(Spergel et al. 2007)  the total mass of a spherical overdensity is expressed as

$$M_T/M_{\odot}=2.12\times 10^{12}\times (R_0/{\rm Mpc})^3. \eqno(10) $$

An important circumstance here is that the estimate of the total mass of a group
corresponds to the scale of $R_0$, which is $\sim3.7$ times larger than its virial
radius.

The analysis of observational data on radial velocities and separations of
galaxies in the vicinity of the Local Group and other nearby groups was done
by different authors. A summary for six groups is presented in Table 4. The
columns of the table list:  (1) name of the group; (2) logarithm of
the orbital mass of the group in the units of solar mass  and its error;
(3) radius of the  zero velocity sphere (in Mpc) and its error; (4)
logarithm of the total mass of the group, determined   by eq. (10)
and its  error; (5) the difference of the total and orbital mass estimates;
(6) the reference to the source of data on $R_0$.

In general, the estimates of mass by two independent methods
agree  with each other quite well.  However,  a moderate systematic difference
of mass estimates in favour of the orbital masses is noteworthy.
For six groups the mean difference amounts to
$\langle\Delta\lg(M_T/M_{orb})\rangle=-0.20\pm0.05$. This paradoxical result
lying in the fact that the estimates of the total mass of the groups on
the scale of  $R_0\sim3.7 R_v$  are lower than the orbital (as well as the projected)
mass estimates on the scale of the virial radius $R_v$ can have a simple
interpretation. Chernin et al. (2013) noted that the estimate of the total
mass of a group includes two components:
$M_T = M_M + M_{DE}$, where $M_M$ is the mass of dark and baryonic matter,
and $M_{DE}$ is the mass, negative in magnitude,  determined by the dark
energy with the density of  $\rho_{DE}$:
$$ M_{DE}= (8\pi/3)\times\rho_{DE}\times R^3.$$

On the scale of   $R_v$ the contribution of this component in the group mass
is small, not exceeding 1\%. But in the sphere of $R_0$  radius, the
role of this ``mass defect'' becomes significant. In the standard
$\Lambda$CDM  model with $\Omega_m =0.24$ the contribution of dark energy is

$$(M_{DE}/M_{\odot})= -0.85\times 10^{12} \times (R_0/ {\rm Mpc})^3, \eqno(11)$$
i.e. about 40\% of the value determined by  eq. (10). A correction
to the total mass by a factor of 1.4 can almost completely eliminate the
observed discrepancy between the group mass estimates at different scales.

In turn, such an agreement of mass estimates by the internal and external
motions after the correction for the dark energy component can be interpreted
as another empirical evidence for the existence
of the dark energy itself appearing in the dynamics of nearby groups.

\section{Concluding remarks}
The high-density data on the distances and radial velocities of  $\sim800$
most nearby galaxies from the  UNGC catalog provides an unique opportunity
to investigate  the distribution of light and dark matter in the Local Volume
of  $\sim10$ Mpc radius  in outstanding detail. The analysis of these data
shows that about a half of the population of the Local Volume is concentrated
in the rooms,   dominated by the gravitational influence of only 15
most massive galaxies. Ranking the galaxies by the magnitude of tidal
force  allows to group small  galaxies in suites around their   Main Disturbers.
Assuming the Keplerian motions of the companions around the
central galaxy with a typical orbit eccentricity of   $e^2=1/2$,
we have determined the  orbital masses of the main galaxies in the Local
Volume, as well as the total mass of less populated suites. Wherein,
we did not use any restrictions on the  radial velocity of
companions relative to their main galaxy in the suites.

For the mass of dark halo around the MW and around M31,
we have obtained the values of (1.35$\pm$0.47) and (1.76$\pm$0.33) in
the units of $10^{12} M_{\odot}$, respectively. Analyzing the mass estimates
of these galaxies, made by various authors and via different methods,
Shull (2014) has concluded that the virial masses of MW and M31
amount to   ($1.6\pm0.4)$ and $(1.8\pm0.5)(\times10^{12}M_{\odot}$),  what
is in a remarkable agreement with our estimates. The total mass of the
Local Group from our data is $(3.1\pm0.6)(\times10^{12}M_{\odot}$). This
estimate is consistent with the (MW + M31) mass estimate by Partridge et al.
(2013) and Gonzalez et al. (2013) obtained based on the timing argument.

Within the Local Volume, there are 18 groups identified with the suites,
for which MK11 estimated the virial (projected)
masses $M_p$. On average, the agreement between
the orbital and projected  mass estimates for the suites and groups proves
to be quite satisfactory.
The typical ratio of both the orbital or the projected mass-to-sum of
stellar masses of  galaxies forming the group
amounts to $M_{orb}/\Sigma M_*\simeq 30$.

Among the smallest suites in the Local Volume there are 12 isolated dwarf
pairs, where each galaxy with a characteristic stellar mass of
$\sim10^8 M_{\odot}$
is the MD for the second component.  The average ratio of the orbital mass-
to-sum of stellar masses for them,
$\langle M_{orb}/\Sigma M_*\rangle=237\pm172$,  looks a little more than that
for the suites around luminous galaxies.
However, taking account of a significant gas component in these small binary
systems leads to the baryonic ratio of
$\langle M_{orb}/\Sigma(M_*+M_{gas})\rangle = 78\pm56$,
close to the typical one of the galaxies with normal luminosities.

The distortion in the Hubble flow, observed around six most nearby groups
allows us to determine their total masses. Independent estimates of total
masses via the radius of zero velocity sphere $R_0$ are slightly lower
than the orbital and virial values. This difference may be due to the local
effect of dark energy, which affects the kinematics of the galaxy groups,
especially scattered ones.

The data we have obtained on the orbital masses of suites/groups, summed
over the Local Volume of the 8 Mpc radius, yield the ratio of dark-to-luminous
matter of  $\Sigma M_{orb}/\Sigma M_*\simeq 30$, which corresponds to
the mean local density of  $\Omega_m\simeq 0.09$.  It seems difficult to
indicate the precise error of this value, because the error is rather dominated
by systematic effects than by random statistics. The present result is in line
with the measurement $\Omega_m= 0.08\pm0.02$, derived by MK11 within a volume
of the Local Supercluster ($D < 50$ Mpc) using an independent approach to find
galaxy groups. Therefore, a threefold divergence between the local
and global values of  $\Omega_m$, noted by many authors, remains to be
an unsolved mystery of the near-field cosmology.

{\bf Acknowledgments}

 We thank the anonymous referee for thorough reading the manuscript
and valuable comments.
This work was supported by the grant of the Russian Foundation for Basic
Research   13--02--90407 Ukr-f-a and the grant of the Ukraine F53.2/15.
IK acknowledge the support for proposals GO 12877 and 12878
provided by NASA through grants from the Space Telescope Science Institute.

{\bf References}

 Bahcall N.A., Kulier A., 2014, MNRAS, 439, 2505

 Bahcall N.A., Cen R., Dave R., et al. 2000, ApJ, 541, 1

 Bahcall J.N., Tremaine S., 1981, ApJ, 244, 805

 Barber C., Starkenburg E., Navarro J.F.,et al. 2014, MNRAS, 437, 959

 Bell E.F., McIntosh D.H., Katz N., Weinberg M.D., 2003, ApJS, 149, 289

 Beloborodov A.M., Levin Y., 2004, ApJ, 613, 224

 Belokurov, V., Zucker, D. B., Evans, N. W., et al. 2006, ApJL, 647, L111


 Bond J.R., Kofman L., Pogosyan D., 1996, Nature, 380, 603


 Cautun M., van de Weygaert R., Jones B.J.T., Frenk C.S., 2014, arXiv:1401.7866

 Chernin A.D., Bisnovatyi-Kogan G.S., Teerikorpi P., et al. 2013, A\&A, 553, 101

 Crook A.C., Huchra J.P., Martimbeau N. et al., 2007, ApJ, 655, 790

 Einasto J., Hutsi G., Saar E., et al. 2011, A\&A, 531A, 75


 Fukugita, M., \& Peebles, P. J. E. 2004, ApJ, 616, 643

 Gonzalez R.E., Kravsov A.V., Gnedin N.Y., 2013, arXiv:1312.2587

 Heisler J., Tremaine S., Bahcall J. N., 1985, ApJ, 298, 8

 Huchra J.P., Geller M.J., 1982, ApJ, 257, 423

 Ibata, R., Martin, N. F., Irwin, M., et al. 2007, ApJ, 671, 1591

 Jones D. H., Peterson B. A., Colless M., Saunders W., 2006, MNRAS, 369, 25

 Karachentsev I.D., Kaisina E.I., Makarov D.I., 2014a, AJ, 147, 13

 Karachentsev I.D., Tully R.B., Wu P.F., Shaya E.J., Dolphin A.E., 2014b, ApJ, 782, 4

 Karachentsev I.D., Makarov D., Kaisina E., 2013, AJ, 145, 101


 Karachentsev I.D., Kashibadze O.G., Makarov D.I., Tully R.B., 2009, MNRAS, 393, 1265

 Karachentsev I.D., Makarov D.I., 2008, Astrophys. Bulletin, 63, 299

 Karachentsev I.D., Tully R.B., Dolphin A.E., et al. 2007, AJ, 133, 504

 Karachentsev I.D., Kashibadze O.G., 2006, Astrophysics, 49, 3

 Karachentsev I.D., 2005, AJ, 129, 178

 Karachentsev I.D., Sharina M.E., Dolphin A.E., Grebel E.K., 2003a, A\&A, 408, 111

 Karachentsev I.D., Grebel E.K., Sharina M.E., et al. 2003b, A\&A, 404,93

 Karachentsev I.D., 1966, Astrophysics, 2, 159

 Knebe A., Libeskind N.I., Doumler T., et al. 2011, MNRAS, 417L, 56

 Libeskind N.I., Hoffman Y., Forero-Romero J., et al, 2013, MNRAS, 428, 2489

 Libeskind N.I., Yepes G., Knebe A., et al, 2010, MNRAS, 401, 1889

 Lokas E.L., Mamon G.A., 2001, MNRAS, 321, 155

 Lynden-Bell D., 1981, Observatory  101, 111

 Magtesian A., 1988, Astrofizika, 28, 150

 Makarov D.I., Makarova L.N., Uklein R.I., 2013, AstBu, 68, 125

 Makarov D.I., Uklein R.I., 2012, Astr. Bull., 67, 135

 Makarov D.I., Karachentsev I.D., 2011, MNRAS, 412, 2498 (=MK11)

 Martin, N. F., McConnachie, A. W., Irwin, M., et al. 2009, ApJ, 705, 758

 Mateo M., 1998, ARA\&A, 36, 435


 Moster B.P., Somerville R.S., Maulbetsch C., et al. 2010, ApJ, 710, 903


 Partridge C., Lahav O., Hoffman Y., 2013, MNRAS, 436L, 45



 Rood H.J., Rothman V.C.A., Turnrose B.E., 1970, ApJ, 162, 411

 Sandage A., 1986, ApJ 307, 1

 Shandarin S.F., Sheth J.V., Sahni V., 2004, MNRAS, 353, 162

 Shull J.M., 2014, ApJ, 784, 142


 Spergel D.N., et al. 2007, ApJS, 170, 377

 Tully R.B., 1987, ApJ, 321, 280

 Vennik J., 1984, Tartu Astron. Obs. Publ., 73, 1



 Willman, B., Dalcanton, J. J., Martinez-Delgado, D., et al. 2005, ApJL, 626, L85

 Wojtak R., 2013, A \& A, 559, 89

 Zavala J., Jing Y.P., Faltenbacher A., et al, 2009, ApJ, 700, 1779
\renewcommand{\baselinestretch}{0.6}
\begin{table}
\caption{Basic properties of the nearby galaxy suites}
\begin{tabular}{lrrrcrr}
\hline
Main galaxy&   $N_v$   &
\multicolumn{1}{c}{$\langle R_p\rangle$}  &
\multicolumn{1}{c}{$\langle|dV|\rangle$}    &
\multicolumn{1}{c}{$M*_{MD}$}            &
\multicolumn{2}{c}{ $\langle M_{orb}\rangle$}  \\ \hline
(1) & (2) & (3) & (4) & (5) &\multicolumn{2}{c}{(6)}\\ \hline

Milky Way      & 27     &  121     &  84      &  5.0      &   1.35&$\pm$ 0.47 \\
M31            & 39     &  198    &   93      &  5.4      &     1.76&$\pm$ 0.33\\
MW + M31      & 66     &  167    &   89      &  5.2      &   1.56&\\
& & & & & &\\
M81             & 26     &  219    &  116      &  8.5       &  4.89&$\pm$ 1.41\\
N5128           & 15     &  343    &  110      &  8.1       &  6.71&$\pm$ 2.09\\
N4594            &  6     &  577    &  153      & 20.0       & 28.47&$\pm$ 17.80\\
N3368           & 20     &  408    &  150      &  6.8       & 17.00&$\pm$ 4.30\\
N4258           & 11     &  316     &  96      &  8.7       &  3.16&$\pm$ 1.01\\
N4736          & 14     &  515     &  50      &  4.1       &  2.67&$\pm$ 0.90\\
N5236            &10     &  294     &  57      &  7.2       &  1.06&$\pm$ 0.28\\
N253            & 7     &  500     &  51      & 11.0       &  1.51&$\pm$ 0.59\\
N3115            &  6     &  215     &  82      &  8.9        & 3.43&$\pm$ 2.00\\
M101            &  6     &  167     &  76      &  7.1       &  1.47&$\pm$ 0.67\\
IC342            & 8     &  321     &  66      &  4.0        & 1.81&$\pm$ 0.82\\
N3627            & 7     &  254    &   69      & 10.2        & 1.45&$\pm$ 0.39\\
N6946            & 6     &  163     &  60     &   5.8        & 0.66&$\pm$ 0.34\\
All 13          &142     &  332     &  96      &  8.5        & 6.27&\\
& & & & & & \\
Synth all         & 89     &  188     &  69      &  2.6    &  2.74&$\pm$ 0.77\\
Synth L           & 30      & 352     &  73      &  6.3     & 5.76&$\pm$ 2.09\\
Synth M         & 29     &  156     &  79      &  1.3    & 2.08&$\pm$ 0.68\\
Synth S           &30      &  56      & 55      &  0.18    & 0.34&$\pm$ 0.13\\
\hline
\multicolumn{7}{l}{\parbox{10cm}{The columns contain: (1) name of the suite/group by its main galaxy,
(2) the number of physical ($\Theta_1\geq0$) members of the group with
measured radial velocities,
(3)  the average projection separation of the companions from the main galaxy
(kpc),
(4)  the mean absolute value of the  radial velocity difference of the
companions relative to the main galaxy
(km/s),
(5) the main galaxy stellar mass in the units of  $10^{10}M_{\odot}$,
(6) the  value of orbital mass of the group
(suite) with the standard error  in  units of
$10^{12}M_{\odot}$.
The location of suites in Table 1 corresponds to their breakdown in the
three panels of Fig. 1:
the first lines contain  the data for the MW and M31 groups,
followed by  the characteristics of 13 other most populated groups of the
Local Volume, and   the end of the table shows the average parameters of a
composite (synthetic) suite.}}
\end{tabular}
\end{table}

\renewcommand{\baselinestretch}{0.3}
 \begin{table}
 \caption{ The nearby suites common with the MK(2011) groups}
\begin{tabular}{lrrrrcccrrrrr}\\ \hline
 Group &  Nv    &$D_H$   &$\sigma_V$  & $R_h$ & $\log M_*$ & $\log M_p$ & $\log M_p/M_*$ & $T$ & $\Delta M_{12}$ & $\Theta_1$  &$\Theta_5$& $\Theta_J$\\

 Suite &  Nv  &   $D_{MD}$   &$\Delta V$  &   $R_{12}$ & $\log M_*$  & $\log M_{orb}$ &$\log M_{orb}/M_*$
 &   &      &       &      &       \\
\hline
(1) & (2)& (3)& (4)& (5)& (6)& (7)& (8) &(9) &(10& (11)& (12)& (13)\\ \hline
N253   &   6  &  5.1  &  87  &  275  & 11.24& 12.87 &  1.63  &   5 & 3.66 &-0.3 &0.2 & 0.7
  \\
       &   8  &  3.9  &  64  &  500  & 11.07  & 12.18   & 1.11
    &   &      &       &      &        \\
  & & & & & & & & & & & & \\
    N628   &   6  & 11.4  &  46  &  171  & 10.71  & 12.18   & 1.47      &5  & 4.96 & -0.4  &-0.2  &-0.5
 \\
       &   5  &  7.3  &  82  &  230  & 10.32  & 12.20   & 1.88
    &   &      &       &      &        \\
  & & & & & & & & & & & & \\
    N672   &   5  &  7.7  &  41  &   74  &  9.78  & 11.39   & 1.61      &5  & 1.78 &  3.8  & 3.8  & 0.2
\\
       &   4  &  7.2  &  67  &  105  &  9.87  & 11.66   & 1.79
    &   &      &       &      &        \\
  & & & & & & & & & & & & \\
    N891   &  18  & 10.6  &  60  &  197  & 11.30  & 12.64   & 1.34      &3  & 0.30 & -0.9  &-0.5  &-0.1
 \\
       &   4  &  9.8  &  35  &  607  & 10.96  & 11.90   & 0.94
    &   &      &       &      &        \\
  & & & & & & & & & & & & \\
    N2903  &   4  &  5.7  &  31  &   69  & 10.42  & 11.62   & 1.20      &4  & 5.64 &  1.6  & 1.6  &-0.8
  \\
       &   5  &  8.9  &  45  &  197  & 10.82  & 11.68   & 0.86
    &   &      &       &      &         \\
  & & & & & & & & & & & & \\
    M81    &  30  &  2.6  & 138  &  102  & 10.86  & 12.59   & 1.73      &3  & 0.81 &  2.5  & 2.6  & 1.5
   \\
       &  27  &  3.6  & 133  &  219  & 11.11  & 12.69   & 1.58
    &   &      &       &      &         \\
  & & & & & & & & & & & & \\
    N3115  &   5  &  6.0  &  58  &  119  & 10.53  & 12.29   & 1.76    &-1  & 4.17 &  2.2  & 2.5  & 0.2
   \\
       &   7  &  9.7  & 113  &  215  & 10.96  & 12.54   & 1.57
    &   &      &       &      &        \\
  & & & & & & & & & & & & \\
    N3379  &  27  & 10.2  & 233  &  179  & 11.47  & 13.23   & 1.76      &2  & 0.05 &  1.2  & 1.5  & 2.1
   \\
   N3368  &  21  & 10.4  & 175  &  408  & 11.13  & 13.23   & 2.10
    &   &      &       &      &         \\
  & & & & & & & & & & & & \\
    N3627  &  16  & 10.0  & 154  &  192  & 11.43  & 13.05   & 1.62      &4  & 0.19 &  1.1  & 1.3  & 2.0
  \\
       &   8  & 10.3  &  78  &  254  & 11.11  & 12.16   & 1.05
    &   &      &       &      &         \\
  & & & & & & & & & & & & \\
    N4258  &  15  &  7.6  &  80  &  254  & 10.97  & 12.45   & 1.48      &4  & 2.34 &  1.2  & 1.4  & 1.0
   \\
       &  12  &  7.8  & 127  &  316  & 10.97  & 12.50   & 1.53
    &   &      &       &      &         \\
  & & & & & & & & & & & & \\
    N4594  &  11  & 11.7  &  61  &  597  & 11.53  & 12.90   & 1.37      &1  & 2.98 &  2.7  & 2.8  &-0.4
   \\
       &   7  &  9.3  & 188  &  577  & 11.30  & 13.45   & 2.15
    &   &      &       &      &         \\
  & & & & & & & & & & & & \\
    N4631  &  28  &  8.7  &  90  &  243  & 11.12  & 12.98   & 1.86      &7  & 0.25 &  1.8  & 1.8  & 1.0
   \\
       &   5  &  7.4  & 191  &  338  & 10.54  & 13.24   & 2.70
    &   &      &       &      &         \\
  & & & & & & & & & & & & \\
    N4736  &   5  &  4.8  &  16  &  338  & 10.64  & 11.34   & 0.70      &2  & 5.49 & -0.6  &-0.1  & 0.8
   \\
       &  15  &  4.7  &  66  &  515  & 10.72  & 12.43   & 1.70
    &   &      &       &      &        \\
  & & & & & & & & & & & & \\
    N5128  &  15  &  4.1  &  94  &  402  & 11.21  & 12.52   & 1.31     &-2  & 0.52 &  0.7  & 1.0  & 1.6
   \\
       &  16  &  3.8  & 137  &  343  & 11.17  & 12.83   & 1.66
    &   &      &       &      &         \\
  & & & & & & & & & & & & \\
    N5194  &   9  &  7.9  &  84  &  182  & 11.29  & 12.93   & 1.64      &4  & 0.12 &  0.0  & 0.4  & 1.3
   \\
       &   4  &  8.4  &  53  &  167  & 11.12  & 11.78   & 0.66
    &   &      &       &      &         \\
  & & & & & & & & & & & & \\
    N5236  &  12  &  4.4  &  77  &  149  & 10.78  & 12.29   & 1.51      &5  & 3.63 & -0.5  & 0.0  & 0.0
  \\
       &  11  &  4.9  &  61  &  294  & 10.87  & 12.02   & 1.15
    &   &      &       &      &         \\
  & & & & & & & & & & & & \\
    M101   &   6  &  5.2  &  61  &  150  & 10.56  & 12.05   & 1.49      &6  & 3.97 &  0.3  & 0.5  & 0.2
   \\
       &   7  &  7.4  &  81  &  167  & 10.86  & 12.17   & 1.30
    &   &      &       &      &         \\
  & & & & & & & & & & & & \\
    N6744  &   9  & 10.3  &  78  &  229  & 11.12  & 12.59   & 1.47      &4  & 1.11 &  2.0  & 2.0  & 1.1
   \\
       &   4  &  8.3  &  90  &  401  & 10.94  & 12.70   & 1.75
    &   &      &       &      &         \\
\hline \\
 Mean  &  13  &  7.4  &  83  &  218  & 10.94  &12.44  & 1.50  &   3 &2.33&  1.0 &1.2 & 0.7  \\
       &   9  &  7.4  &  99  &  325  & 10.88  & 12.41   & 1.53
    &   &      &       &      &         \\
\hline       \\
\multicolumn{12}{l}{\parbox{16cm}{The columns contain:
(1)  name of the main galaxy of the group/suite;
(2)  the number of galaxies in the group/suite with measured radial
velocities;
(3)  the distance to the group (Mpc), determined by the mean radial
velocity  of the group members relative to the Local Group centroid
at  $H_0=73$ km/s/Mpc, and the individual distance of the
principal galaxy of the suite;
(4)  dispersion of radial velocities in the group and the mean-square
difference of the companion velocities  relative to
the main galaxy (km/s);
(5)  the mean harmonic  radius of the group and the
mean projection separation of the companions from the main galaxy
(kpc);
(6)  logarithm of the total stellar mass of the group or a suite (in
$M_{\odot}$);
(7) logarithm of the projected mass of the group and the orbital mass
of the suite (in   $M_{\odot}$);
(8) the ratio of the projected/orbital mass-to-total stellar mass
in the logarithmic scale;
(9) morphological type of the main galaxy on de Vaucouleurs scale;
(10) the difference between the apparent K-magnitudes of the first and
second members of the group;
(11--13)  the tidal indices, characterizing the density of the environment
of the main galaxy of the group: here the $\Theta_1$ index,
determined by equation (6), expresses the contribution of the most
significant  neighbor, the  $\Theta_5$ index
accounts for the effect of five important neighbors, while
the  $\Theta_J$ index corresponds to the logarithm of the stellar density
contrast in a sphere of  1 Mpc radius
around the main galaxy taken with respect to the mean cosmic density.
}}
\end{tabular}
\end{table}

\renewcommand{\baselinestretch}{0.6}
\begin{table}
\caption{Isolated binary dwarfs}
\begin{tabular}{lllrrrrrr} \\

\hline
 &  &&  & &&&\\
Name      &     $D$       & $\Theta_1$  &   $\log M_*$&  $\log M_{HI}$   & $\Delta V$ &  $\sigma_V$ & $R_p$  &$\log M_{orb}$ \\
\hline (1) & (2)& (3)& (4)& (5)& (6)& (7)& (8) &(9) \\ \hline
 &  &&  & &&&\\
N3109     &   1.32    & 0.2   & 8.57   & 8.37     &  44    &  2    & 27 &  10.04\\
Antlia    &   1.32   &  2.3   & 6.47   & 5.92    &         &  1&&\\
          &             &  &&  & &&&\\
Dwing2    &   3.0    &  2.8   & 8.35   & 8.01    &   35    &  2   &  13  & 10.27\\
MB3        &  3.0    &   3.0  &  8.09  &  7.78   &         &   1&&\\
           &            & &&  & &&&\\
KKR 59     &  5.9    &  1.7   & 9.16   &  -       &  11   &   4    & 36  &  9.70\\
KKR 60     &  5.9    &  2.5  &  8.42   &  -       &       &   7&&\\
           &            &  &&  & &&&\\
ESO121-20  &  6.0   & -0.1   & 7.78    &7.94     &  33   &   5    &  6  &  9.85\\
LV0616-57  &  6.0   &   0.6  &  7.07  &  7.36   &        &    4&&\\
           &             & &&  & &&&\\
UGC2716    &   6.4     &  -0.8  &  8.34   & 7.68    &   29  &      1  &  112 &  11.04\\
UGC2684    &   6.5     &  -0.1  &  7.57   & 7.92   &        &     4&&\\
           &            & &&  & & &&\\
KUG1202+28 &  6.7     &  0.1   & 7.70    & -      &    4  &    33   &  16  &  8.49\\
LV1205+281 &  6.7     &   0.5   & 7.37    & -      &      &     14&&\\
           &            & &&  & & &&\\
DDO 64     &   7.1     &  1.6  &  8.04   & 8.24    &   18  &     2   &   4  &  9.21\\
KK 78      &   7.1     &  2.8   & 6.92   & 7.35     &      &    4&&\\
           &            & &&  & & &&\\
DDO161     &   7.3    &   1.5  &  8.91     & 8.99   &      10  &  18  &   40  &  9.67\\
UGCA319    &  7.3    &   2.2  &  8.22     & 7.97   &          &   4&&\\
           &            & &&  & & &&\\
MAPS1206+31&  7.4      & -0.4  &  7.81     &  -   &        4  &  27  &   44  &  8.92\\
LV1207+3133&  7.4      &   0.2  &  7.12    &   -    &       &     4&&\\
           &            & &&  & & &&\\
NGC1156    &  7.8     & -0.3   & 9.31     & 8.82  &     64  &   1    & 80   &11.58\\
LV0300+25  &  7.8    &   1.6  &  7.34     & 6.20 &        &     3&&\\
           &            & &&  & & &&\\
NGC1744    & 10.0     &   0.1 &  9.42     & 9.35   &    90&     2   & 169  & 12.20\\
ESO486-21  & 10.0     & 0.8 &  8.74      &8.47   &       &  18&&\\
           &            & &&  & & &&\\
KK 94      & 10.4     & 2.3 &  7.34     & 7.69   &   12  &   1     & 7   & 9.08\\
LeG 21     & 10.4     & 2.8 &  6.90     & 7.09  &        &   1&&\\
\hline
\multicolumn{9}{l}{\parbox{14cm}{The table columns
 represent  the following data, adopted from the  UNGC catalog: (1)
 names of the components; (2)  their  distances (Mpc); (3)  tidal indices,
characterizing the degree of mutual gravitational influence;
(4) logarithm of stellar
mass ($M_{\odot}$); (5) logarithm of the hydrogen mass  ($M_{\odot}$);
(6)  radial velocity difference of the components   (km/s);
(7)   velocity measurement error  (km/s);
(8) projected separation (kpc);
(9) logarithm of orbital mass  ($M_{\odot}$).}}
\end{tabular}
\end{table}

\clearpage
\renewcommand{\baselinestretch}{0.6}
\begin{table}
\caption{Total masses of nearby groups via internal and external motions}
\begin{tabular}{lccccl}

& & & & & \\
  Group   &    $\log(M_{orb})$  &     $R_0$      &    $\log(M_T)$  &  $\log(M_T/M_{orb})$ &  Reference\\
  \hline (1) & (2)& (3)& (4)& (5)& (6)\\ \hline
& & & & & \\
  \hline
& & & & & \\
MW+M31     &  12.50$\pm$0.08 &  0.98 $\pm$0.03 &  12.30 $\pm$0.05  &  -0.20 $\pm$0.09 &  Karachentsev et al. 2009\\
& & & & & \\
IC342      &  12.26$\pm$0.21 &  0.90 $\pm$0.10 &  12.19 $\pm$0.14 & -0.07 $\pm$0.25 &  Karachentsev et al. 2003a\\
& & & & & \\
M81        & 12.69$\pm$0.13 &  1.05 $\pm$0.07 &  12.39 $\pm$0.09 & -0.30 $\pm$0.16 &  Karachentsev \& Kashibadze 2006\\
& & & & & \\
N5128+N5236 & 12.89$\pm$0.14 &  1.26 $\pm$0.15 &  12.63 $\pm$0.15 & -0.23 $\pm$0.21 &  Karachentsev et al. 2007\\
& & & & & \\
N253        & 12.18$\pm$0.18 &  0.70 $\pm$0.10  & 11.86 $\pm$0.18 & -0.32 $\pm$0.25&   Karachentsev et al. 2003b\\
& & & & & \\
N4736       & 12.43$\pm$0.15 &  1.04 $\pm$0.20 &  12.38 $\pm$0.24 & -0.05 $\pm$0.28 &  Makarov et al. 2013\\
& & & & & \\
\hline
\multicolumn{6}{l}{\parbox{17cm}{The
columns of  table list:  (1) name of the group; (2) logarithm of
the orbital mass of the group in the units of solar mass  and its error;
(3) radius of the  zero velocity sphere (in Mpc) and its error; (4)
logarithm of the total mass of the group, determined   by eq. (10)
and its  error; (5) the difference of the total and orbital mass estimates;
(6) the reference to the source of $R_0$ data.}}
\end{tabular}
\end{table}

\clearpage
\begin{deluxetable}{lccc}
\tablewidth{12cm}

\tablecaption{The tidal indices, $\Theta_1$, projected separation, $R_p$, and radial velocity
differences, $dV$, for members of suites around 15 nearby luminous galaxies.}

\startdata
\hline\hline
 \multicolumn{4}{c}{M81}\\ \hline
 \multicolumn{4}{c}{D=3.63 Mpc, $M_*=8.51\times10^{10} M_{\odot}, M_{orb}=(4.89\pm1.42)\times10^{12} M_{\odot}$}\\
\hline
     Name             & $\Theta_1$        &  $R_p$             & $|dV|$ \\
\hline
  1  HolmIX           & 5.1      &  11          & 88  \\
  2  ClumpI           & 4.2      &  24          & 129 \\
  3  KDG061           & 4.0      &  31          & 256 \\
  4  [CKT2009]d0959+68& 4.0      &  35          & 150 \\
  5  ClumpIII         & 3.9      &  39          & 85  \\
  6  NGC2976          & 2.9      &  88          & 38  \\
  7  MESSIER082       & 2.8      &  39          & 224 \\
  8  KDG064           & 2.7      &  103         & 17  \\
  9  IKN              & 2.5      &  84          & 105 \\
  10 HIJASS J1021+6842& 2.3      &  147         & 83  \\
  11 F8D1             & 2.2      &  119         & 96  \\
  12 KDG063           & 2.0      &  169         & 104 \\
  13 DDO078           & 1.9      &  201         & 87   \\
  14 HolmI            & 1.7      &  157         & 187  \\
  15 KDG073           & 1.4      &  323         & 159  \\
  16 UGC05497         & 1.4      &  333         & 163  \\
  17 HS117            & 1.2      &  191         & 12   \\
  18 BK3N             & 1.2      &  12          & 3     \\
  19 DDO082           & 1.1      &  215         & 103   \\
  20 [CKT2009]d0958+66& 1.0      &  142         & 117   \\
  21 IC2574           & 1.0      &  193         & 79    \\
  22 KDG052           & 0.8      &  511         & 167   \\
  23 DDO053           & 0.8      &  525         & 46    \\
  24 HolmII           & 0.7      &  536         & 207   \\
  25 UGC04483         & 0.6      &  440         & 200   \\
  26 KKH37            & 0.0      &  1030        & 110   \\
		      &          &                 &       \\
  \multicolumn{4}{c}{M31}\\ \hline
 \multicolumn{4}{c}{$D=0.77$ Mpc, $M_*=5.37\times10^{10} M_{\odot}$,    $M_{orb}=(1.76\pm0.33)\times10^{12} M_{\odot}$}\\
\hline
     Name             & $\Theta_1$        &  $R_p$             & $|dV|$ \\
\hline
  1  MESSIER032       & 6.6      &  5           & 93          \\
  2  And IX           & 4.0      &  36          & 77          \\
  3  And XVII         & 3.6      &  43          & 51          \\
  4  NGC0205          & 3.6      &  8           & 76          \\
  5  And I            & 3.4      &  44          & 87          \\
  6  And III          & 3.2      &  67          & 53          \\
  7  And XXVII        & 3.0      &  57          & 232         \\
  8  And XV           & 2.9      &  92          & 50          \\
  9  And XXV          & 2.9      &  81          & 199         \\
  10 And XXVI         & 2.8      &  101         & 50          \\
  11 NGC0147          & 2.8      &  100         & 114         \\
  12 And XI           & 2.7      &  101         & 138         \\
  13 And XII          & 2.6      &  94          & 274         \\
  14 And V            & 2.6      &  109         & 114         \\
  15 And XXIII        & 2.4      &  126         & 23          \\
  16 And XX           & 2.4      &  125         & 153         \\
  17 And XIII         & 2.4      &  115         & 82          \\
  18 And XXX          & 2.3      &  114         & 165         \\
  19 And XXI          & 2.2      &  122         & 44          \\
  20 And X            & 2.2      &  76          & 124         \\
  21 And XIV          & 2.2      &  160         & 211         \\
  22 Bol520           & 2.0      &  115         & 35          \\
  23 And II           & 2.0      &  140         & 69          \\
  24 NGC0185          & 2.0      &  96          & 102         \\
  25 And XXIX         & 2.0      &  188         & 106         \\
  26 And XIX          & 1.9      &  104         & 186         \\
  27 And XXIV         & 1.9      &  111         & 156         \\
  28 MESSIER033       & 1.7      &  203         & 63          \\
  29 Cas dSph         & 1.7      &  223         & 24          \\
  30 IC0010           & 1.6      &  256         & 33          \\
  31 And XVI          & 1.5      &  129         & 114         \\
  32 LGS 3            & 1.5      &  279         & 45          \\
  33 Peg dSph         & 1.4      &  277         & 55          \\
  34 And XXVIII       & 1.1      &  407         & 3           \\
  35 Pegasus          & 0.9      &  463         & 89          \\
  36 IC1613           & 0.7      &  634         & 60          \\
  37 And XVIII        & 0.4      &  112         & 15          \\
  38 Cetus            & 0.3      &  1002        & 55          \\
  39 WLM              & 0.0      &  1209        & 13          \\
		      &          &                 &             \\
\multicolumn{4}{c}{MW}\\ \hline
\multicolumn{4}{c}{$D=0.01$ Mpc,     $M_*=5.00\times10^{10} M_{\odot}$    $M_{orb}=(1.35\pm0.47)\times10^{12} M_{\odot}$}\\
\hline
     Name             & $\Theta_1$        &  $R_p$             & $|dV|$ \\
\hline
  1  Sag dSph         & 5.3      &  5           & 169        \\
  2  Segue 1          & 4.3      &  17          & 111        \\
  3  UMa II           & 3.9      &  21          & 33         \\
  4  BootesII         & 3.8      &  37          & 116        \\
  5  Segue 2          & 3.8      &  22          & 43         \\
  6  Willman1         & 3.7      &  34          & 36         \\
  7  ComaI            & 3.7      &  40          & 82         \\
  8  BootesIII        & 3.6      &  49          & 241        \\
  9  LMC              & 3.5      &  49          & 84         \\
  10 Umin             & 3.2      &  59          & 93         \\
  11 BootesI          & 3.2      &  66          & 106        \\
  12 Draco            & 2.9      &  80          & 101        \\
  13 Sculptor         & 2.7      &  90          & 72         \\
  14 SexDSph          & 2.7      &  85          & 75         \\
  15 Carina           & 2.6      &  99          & 13          \\
  16 UMa I            & 2.5      &  84          & 7          \\
  17 Hercules         & 2.1      &  107         & 145        \\
  18 Fornax           & 2.1      &  137         & 60         \\
  19 LeoIV            & 2.0      &  160         & 10         \\
  20 CVnII            & 2.0      &  160         & 96         \\
  21 LeoV             & 1.8      &  180         & 58         \\
  22 LeoII            & 1.6      &  201         & 32         \\
  23 CvnI             & 1.5      &  220         & 78         \\
  24 LeoI             & 1.4      &  223         & 175        \\
  25 LeoT             & 0.7      &  338         & 57         \\
  26 Phoenix          & 0.7      &  440         & 142        \\
  27 NGC6822          & 0.5      &  257         & 43         \\
		      &          &                 &            \\
\multicolumn{4}{c}{NGC5128}\\ \hline
\multicolumn{4}{c}{$D=3.75$ Mpc,     $M_*=8.13\times10^{10} M_{\odot}$,    $M_{orb}=(6.71\pm2.09)\times10^{12} M_{\odot}$}\\
\hline
     Name             & $\Theta_1$        &  $R_p$             & $|dV|$ \\
\hline
  1  ESO324-024       & 2.9      &  104         & 38      \\
  2  ESO269-066       & 2.0      &  190         & 218     \\
  3  NGC5011C         & 2.0      &  148         & 84      \\
  4  ESO270-017       & 1.8      &  198         & 273     \\
  5  KK196            & 1.6      &  141         & 180     \\
  6  KK211            & 1.6      &  242         & 50      \\
  7  NGC5237          & 1.2      &  146         & 188     \\
  8  ESO325-011       & 1.1      &  248         & 1       \\
  9  NGC5206          & 1.0      &  350         & 24      \\
  10 KK221            & 1.0      &  376         & 42      \\
  11 NGC4945          & 0.9      &  482         & 11      \\
  12 NGC5102          & 0.8      &  422         & 83      \\
  13 ESO383-087       & 0.6      &  549         & 202     \\
  14 PGC051659        & 0.3      &  768         & 133     \\
  15 NGC5253          & 0.3      &  779         & 117     \\
		      &          &                 &         \\
     \multicolumn{4}{c}{NGC4594}\\ \hline
\multicolumn{4}{c}{$D=9.30$  Mpc,    $M_*=19.95\times10^{10} M_{\odot}$,    $M_{orb}=(28.47\pm17.80)\times10^{12} M_{\odot}$}\\
\hline
     Name             & $\Theta_1$        &  $R_p$             & $|dV|$ \\
\hline
  1  SUCD1            & 6.5      &  8           & 215    \\
  2  KKSG30           & 1.2      &  458         & 23     \\
  3  LV J1235-1104    & 0.8      &  194         & 109    \\
  4  MCG-02-33-075    & 0.4      &  757         & 279    \\
  5  DDO148           & 0.2      &  1097        & 276    \\
  6  NGC4597          & 0.0      &  949         & 18     \\
		      &          &                 &        \\
     \multicolumn{4}{c}{NGC3368}\\ \hline
              \multicolumn{4}{c}{$D=10.42$ Mpc,     $M_*=6.76\times10^{10} M_{\odot}$,    $M_{orb}=(17.00\pm4.30)\times10^{12} M_{\odot}$}\\
\hline
     Name             & $\Theta_1$        &  $R_p$             & $|dV|$ \\
\hline
  1  LeG13            & 3.1      &  82          & 22     \\
  2  LeG17            & 2.9      &  92          & 140    \\
  3  FS04             & 1.8      &  231         & 119    \\
  4  UGC05812         & 1.5      &  285         & 117    \\
  5  AGC202456        & 1.5      &  285         & 71     \\
  6  NGC3412          & 1.3      &  343         & 38     \\
  7  LeG05            & 1.3      &  346         & 111    \\
  8  AGC205268        & 1.1      &  390         & 261    \\
  9  NGC3351          & 1.1      &  126         & 117    \\
  10 AGC205445        & 1.1      &  398         & 250    \\
  11 LSBC D640-12     & 1.0      &  419         & 41     \\
  12 LSBC D640-13     & 1.0      &  423         & 100    \\
  13 AGC200499        & 0.9      &  466         & 273    \\
  14 LeG06            & 0.8      &  486         & 123    \\
  15 NGC3299          & 0.8      &  488         & 287    \\
  16 UGC06014         & 0.8      &  491         & 232    \\
  17 AGC202248        & 0.7      &  531         & 280    \\
  18 AGC205156        & 0.3      &  716         & 22     \\
  19 AGC205165        & 0.2      &  771         & 154    \\
  20 LeG03            & 0.2      &  782         & 247    \\
		      &          &                 &        \\
     \multicolumn{4}{c}{NGC4258}\\ \hline             
\multicolumn{4}{c}{$D=7.83$  Mpc,     $M_*=8.71\times10^{10} M_{\odot}$,    $M_{orb}=((3.16\pm1.01)\times10^{12} M_{\odot}$}\\
\hline
     Name             & $\Theta_1$        &  $R_p$             & $|dV|$ \\
\hline
  1  NGC4242          & 1.8      &  233         & 62     \\
  2  NGC4288          & 1.7      &  144         & 82     \\
  3  NGC4248          & 1.1      &  30          & 38     \\
  4  LV J1203+4739    & 0.9      &  372         & 41     \\
  5  KDG101           & 0.7      &  30          & 316    \\
  6  NGC4144          & 0.6      &  239         & 189    \\
  7  KK133            & 0.5      &  536         & 95     \\
  8  UGC07639         & 0.4      &  255         & 56     \\
  9  DDO120           & 0.3      &  211         & 10     \\
  10 MAPS1249+44      & 0.2      &  834         & 66     \\
  11 UGC07827         & 0.1      &  597         & 103    \\
		      &          &                 &        \\
     \multicolumn{4}{c}{NGC4736}\\ \hline
\multicolumn{4}{c}{$D=4.66$ Mpc,     $M_*=4.07\times10^{10} M_{\odot}$,    $M_{orb}=(2.67\pm0.90)\times10^{12} M_{\odot}$}\\
\hline
     Name             & $\Theta_1$        &  $R_p$             & $|dV|$ \\
\hline
  1  IC3687           & 1.4      &  252         & 25     \\
  2  IC4182           & 0.9      &  370         & 5      \\
  3  KK160            & 0.8      &  232         & 6      \\
  4  UGC08215         & 0.5      &  530         & 48     \\
  5  DDO126           & 0.4      &  497         & 122    \\
  6  NGC4449          & 0.3      &  418         & 103    \\
  7  NGC4244          & 0.3      &  592         & 93     \\
  8  DDO168           & 0.3      &  525         & 82     \\
  9  CGCG 189-050     & 0.2      &  486         & 16     \\
  10 DDO169           & 0.2      &  634         & 4      \\
  11 DDO167           & 0.1      &  539         & 122    \\
  12 NGC4395          & 0.1      &  743         & 44     \\
  13 DDO169NW         & 0.0      &  635         & 24     \\
  14 MCG +06-27-017   & 0.0      &  759         & 11     \\
		      &          &                 &        \\
     \multicolumn{4}{c}{NGC5236}\\ \hline
            \multicolumn{4}{c}{ $D=4.92$ Mpc,     $M_*=7.24\times10^{10} M_{\odot}$,    $M_{orb}=((1.06\pm0.28)\times10^{12} M_{\odot}$}\\
\hline
     Name             & $\Theta_1$        &  $R_p$             & $|dV|$ \\
\hline
  1  IC4247           & 2.0      &  195         & 107    \\
  2  UGCA365          & 1.3      &  55          & 60     \\
  3  ESO444-084       & 1.3      &  157         & 73     \\
  4  KK200            & 1.2      &  249         & 36     \\
  5  NGC5264          & 1.1      &  86          & 38     \\
  6  KK195            & 0.9      &  326         & 38     \\
  7  IC4316           & 0.7      &  104         & 62     \\
  8  HIDEEP J1337-33  & 0.5      &  301         & 64     \\
  9  ESO384-016       & 0.4      &  596         & 43     \\
  10 HIPASS J1337-39 & 0.1      &  870         & 49     \\
		      &          &                 &        \\
     \multicolumn{4}{c}{NGC0253}\\ \hline            
 \multicolumn{4}{c}{$D=3.94$ Mpc,     $M_*=10.96\times10^{10} M_{\odot}$,    $M_{orb}=(1.51\pm0.59)\times10^{12} M_{\odot}$}\\
\hline 
    Name             & $\Theta_1$        &  $R_p$             & $|dV|$ \\
\hline
  1  NGC0247          & 1.2      &  312         & 60     \\
  2  ESO540-032       & 0.7      &  374         & 9      \\
  3  DDO006           & 0.6      &  297         & 68     \\
  4  KDG002           & 0.5      &  500         & 14     \\
  5  NGC7793          & 0.2      &  916         & 26     \\
  6  DDO226           & 0.1      &  221         & 133    \\
  7  ESO349-031       & 0.0      &  880         & 46     \\
		      &          &                 &        \\
     \multicolumn{4}{c}{NGC3115}\\ \hline
\multicolumn{4}{c}{$D=9.68$ Mpc,     $M_*=8.91\times10^{10} M_{\odot}$,    $M_{orb}=(3.43\pm2.00)\times10^{12} M_{\odot}$}\\
\hline
     Name             & $\Theta_1$        &  $R_p$             & $|dV|$ \\
\hline
  1  KDG065           & 4.9      &  15          & 40     \\
  2  KKSG18           & 3.9      &  48          & 17     \\
  3  KKSG17           & 2.3      &  175         & 236    \\
  4  MCG -01-26-009   & 1.8      &  254         & 71     \\
  5  UGCA193          & 1.5      &  309         & 12     \\
  6  KKSG15           & 0.9      &  488         & 115    \\
		      &          &                 &        \\
     \multicolumn{4}{c}{M101}\\ \hline
 \multicolumn{4}{c}{$D=7.38$ Mpc,     $M_*=7.08\times10^{10} M_{\odot}$,    $M_{orb}=(1.47\pm0.67)\times10^{12} M_{\odot}$}\\
\hline
     Name             & $\Theta_1$        &  $R_p$             & $|dV|$ \\
\hline
  1  GBT 1355+5439    & 2.2      &  161         & 33     \\
  2  NGC5474          & 2.0      &  95          & 46     \\
  3  HolmIV           & 1.8      &  170         & 106    \\
  4  NGC5477          & 1.4      &  46          & 73     \\
  5  KKH87            & 0.9      &  414         & 95     \\
  6  UGC08882         & 0.0      &  117         & 104    \\
		      &          &                 &        \\
     \multicolumn{4}{c}{IC342}\\ \hline 
    \multicolumn{4}{c}{$D=3.28$  Mpc,     $M_*=3.98\times10^{10} M_{\odot}$,    $M_{orb}=(1.81\pm0.82)\times10^{12} M_{\odot}$}\\
\hline
     Name             & $\Theta_1$        &  $R_p$             & $|dV|$ \\
\hline
  1  KK35             & 2.4      &  16          & 95      \\
  2  UGCA086          & 1.1      &  90          & 36      \\
  3  NGC1560          & 1.0      &  313         & 74      \\
  4  CamB             & 0.9      &  366         & 23      \\
  5  NGC1569          & 0.9      &  313         & 138     \\
  6  Cas1             & 0.5      &  530         & 40      \\
  7  UGCA105          & 0.3      &  612         & 37      \\
  8  CamA             & 0.0      &  327         & 88      \\
		      &          &                 &         \\
      \multicolumn{4}{c}{NGC3627}     \\ \hline
    \multicolumn{4}{c}{$D=10.28$ Mpc,     $M_*=10.23\times10^{10} M_{\odot}$,    $M_{orb}=(1.45\pm0.39)\times10^{12} M_{\odot}$}\\
\hline
     Name             & $\Theta_1$        &  $R_p$             & $|dV|$ \\
\hline
  1  AGC213436        & 3.1      &  94          & 88      \\
  2  IC2684           & 2.6      &  143         & 128     \\
  3  IC2791           & 2.5      &  149         & 49      \\
  4  IC2787           & 2.3      &  176         & 3       \\
  5  AGC215354        & 1.7      &  237         & 80      \\
  6  NGC3593          & 0.8      &  249         & 87      \\
  7  CGCG 066-109     & 0.4      &  728         & 50      \\
		      &          &                 &         \\
    \multicolumn{4}{c}{NGC6946}\\ \hline
  \multicolumn{4}{c}{$D=5.89 $ Mpc,  $M_*=5.75\times10^{10} M_{\odot}$,    $M_{orb}=((0.66\pm0.34)\times10^{12} M_{\odot}$}\\
\hline
     Name             & $\Theta_1$        &  $R_p$             & $|dV|$ \\
\hline
  1  KK251            & 3.5      &  59          & 78      \\
  2  UGC11583         & 3.3      &  65          & 74      \\
  3  KK252            & 3.1      &  80          & 86      \\
  4  KKR55            & 2.4      &  136         & 18      \\
  5  KKR56            & 1.7      &  238         & 91      \\
  6  Cepheus1         & 0.9      &  401         & 13      \\
\hline
\enddata
\end{deluxetable}

\begin{figure}
\epsscale{1.0}
\plotone{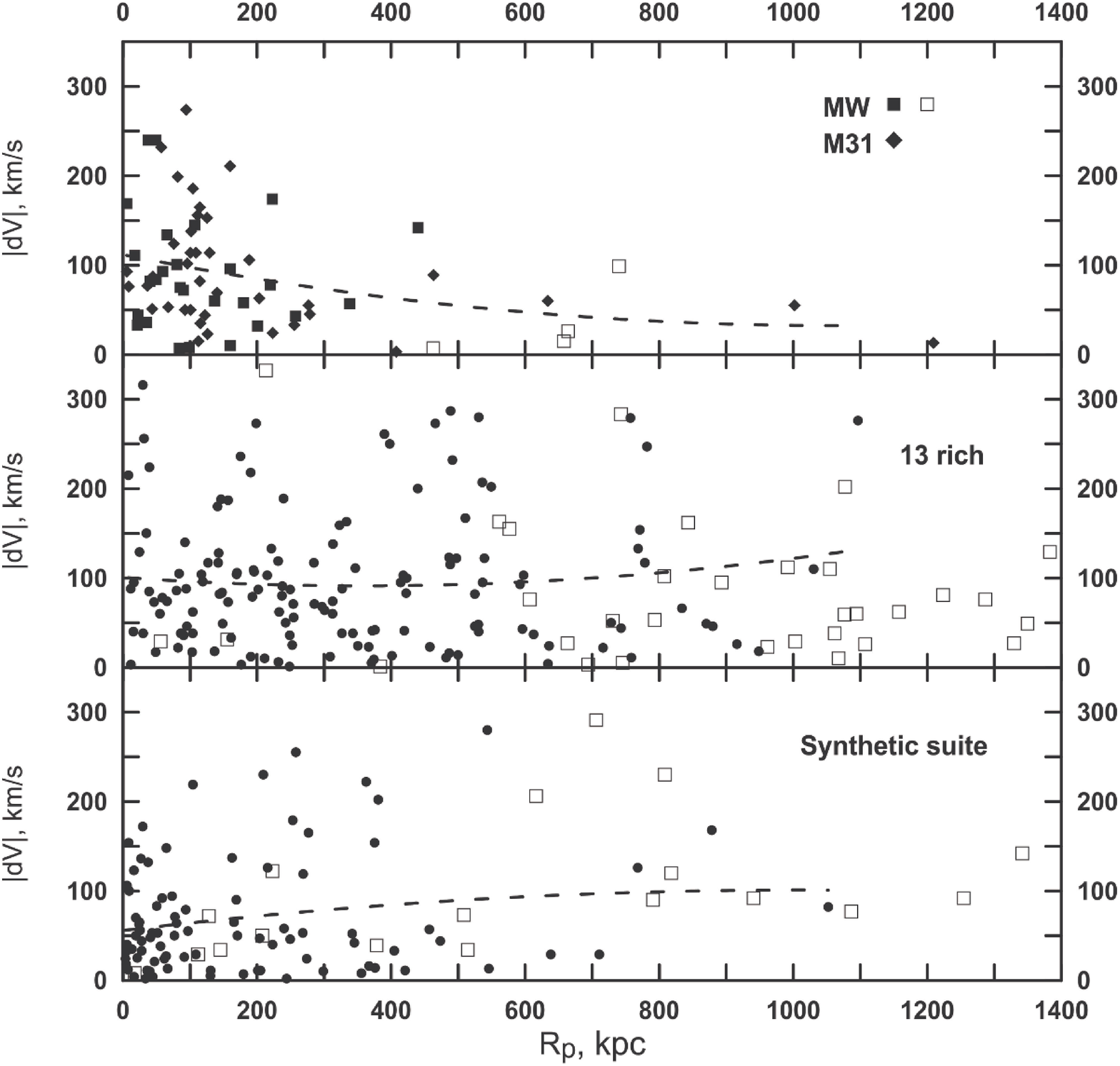}
\caption{Line-of-sight velocity of the suite members relative to the
	   main galaxy as a function of their projected linear separation.
	   The upper panel corresponds to 70 companions of the MW
	   (squares) and M31 (diamonds). The middle panel indicates data on
	   174 galaxies in 13 the most populated nearby suites. The bottom
	   panel presents a synthetic suite formed of 107 companions
	   around other smaller Main Disturbers. Marginal members of
	   the suites with $\Theta_1$ = [-0.5 - 0.0] are depicted by open
	   squares. The dashed lines trace quadratic regressions. }
\end{figure}
\begin{figure}
\epsscale{1.0}
\plotone{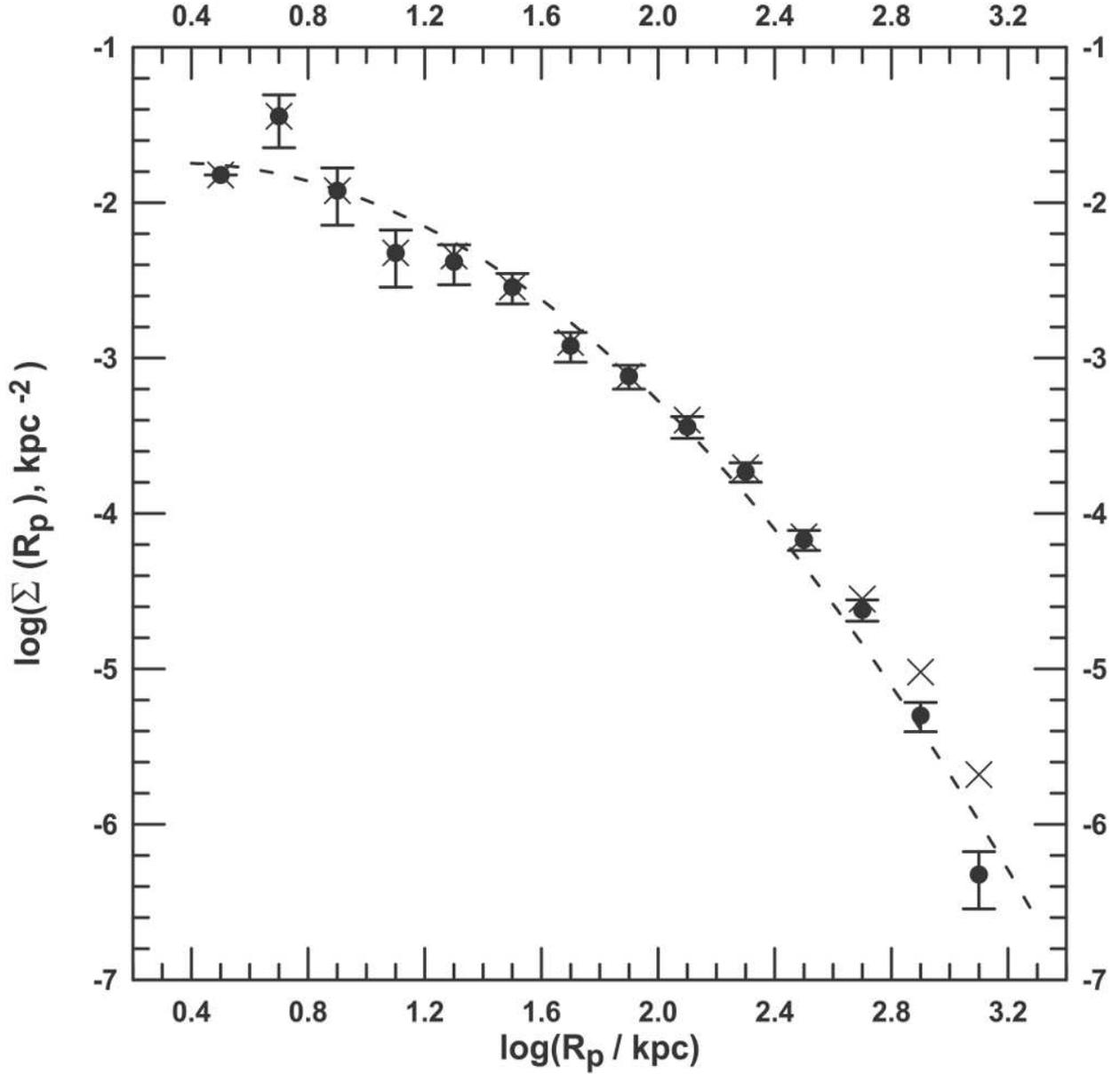}
\caption{The radial profile of the surface number density of companions 
           in the synthetic suite of the Local Volume. The solid circles
	   represent the physical suite members with $\Theta_1 > 0$,
	   and the crosses also account for the marginal members
	   with $\Theta_1= [-0.5 - 0.0]$. The dashed line fits the
	   quadratic regression for the physical companions.}
\end{figure}

\begin{figure}
\epsscale{1.0}
\plotone{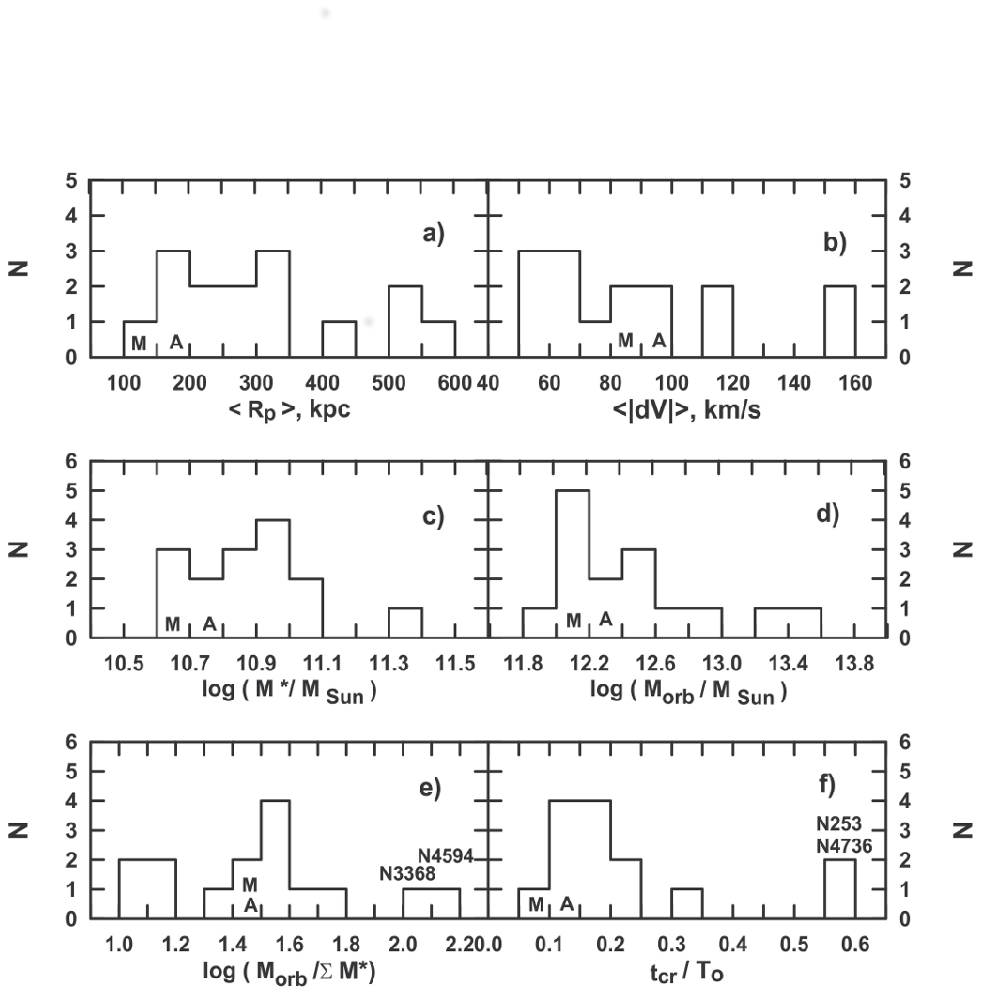}
\caption{The distributions of the 15 richest nearby suites according to:
	   a) the mean projected separation of physical companions,
	   b) radial velocity dispersion, c) logarithm of stellar
	   mass of the main galaxy, d) logarithm of the mean orbital mass
	   estimate, e) the mean orbital-to-stellar mass ratio, f) the
	   mean crossing time for the components in units of the global cosmic
	   time $T_0$. The Milky Way suite and the Andromeda (M31) suite
	   are depicted by "M" and "A", respectively.}
\end{figure}
\begin{figure}
\epsscale{1.0}
\plotone{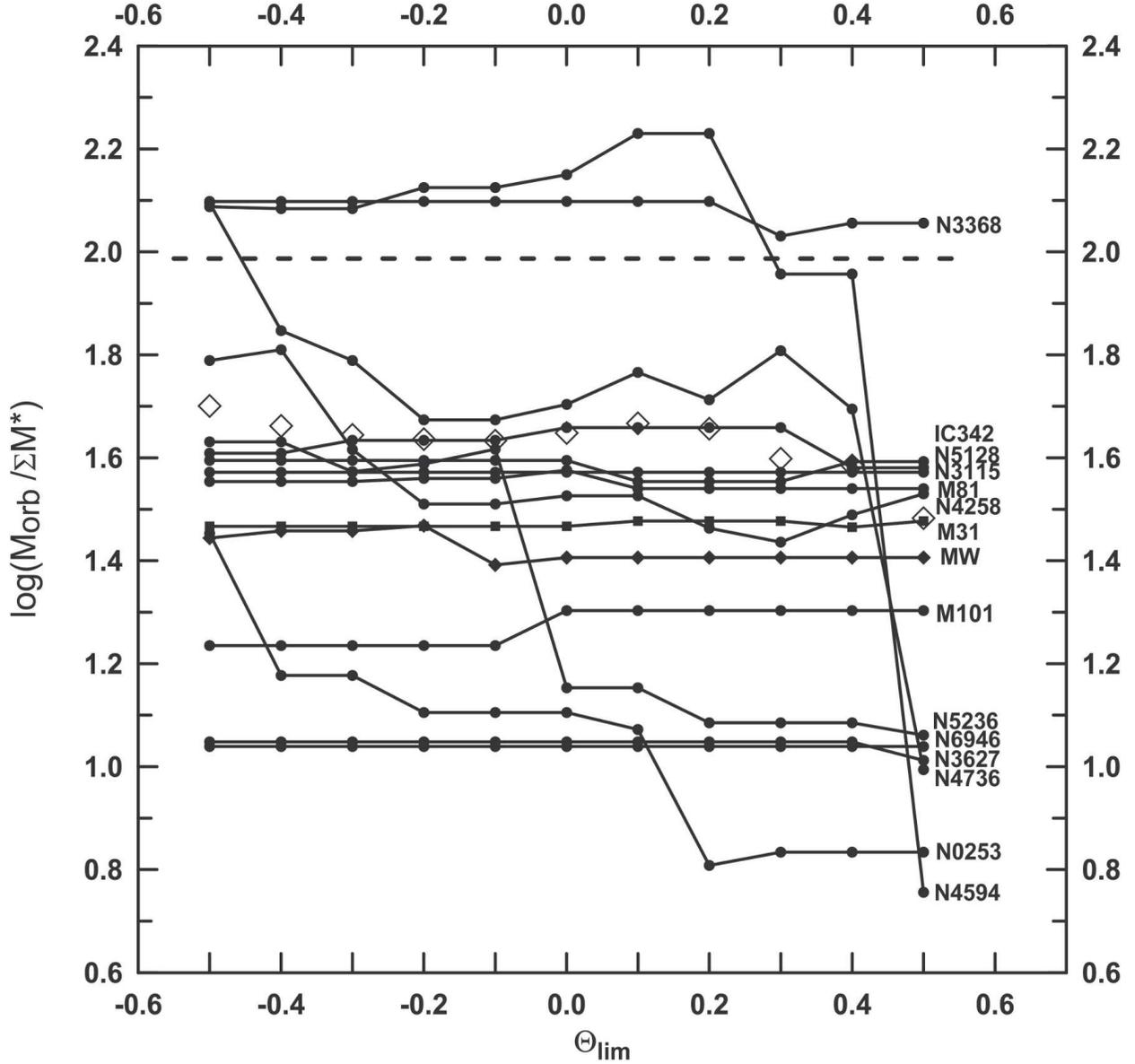}
\caption{The orbital mass-to-sum of stellar mass ratio for 15
           richest suites as a function of cutoff on $\Theta_1$. The suites
	   of the MW and M31 are marked by solid diamonds and squares,
	   respectively. Large open diamonds show the weighted average orbital
	   mass-to-stellar mass ratio for the sample of 15 suites. The dashed
	   horizontal line indicates the $M_{DM}/M_* = 97$ ratio corresponding
           to $\Omega_m = 0.28$.}
 \end{figure}
\begin{figure}
\epsscale{0.8}
\plotone{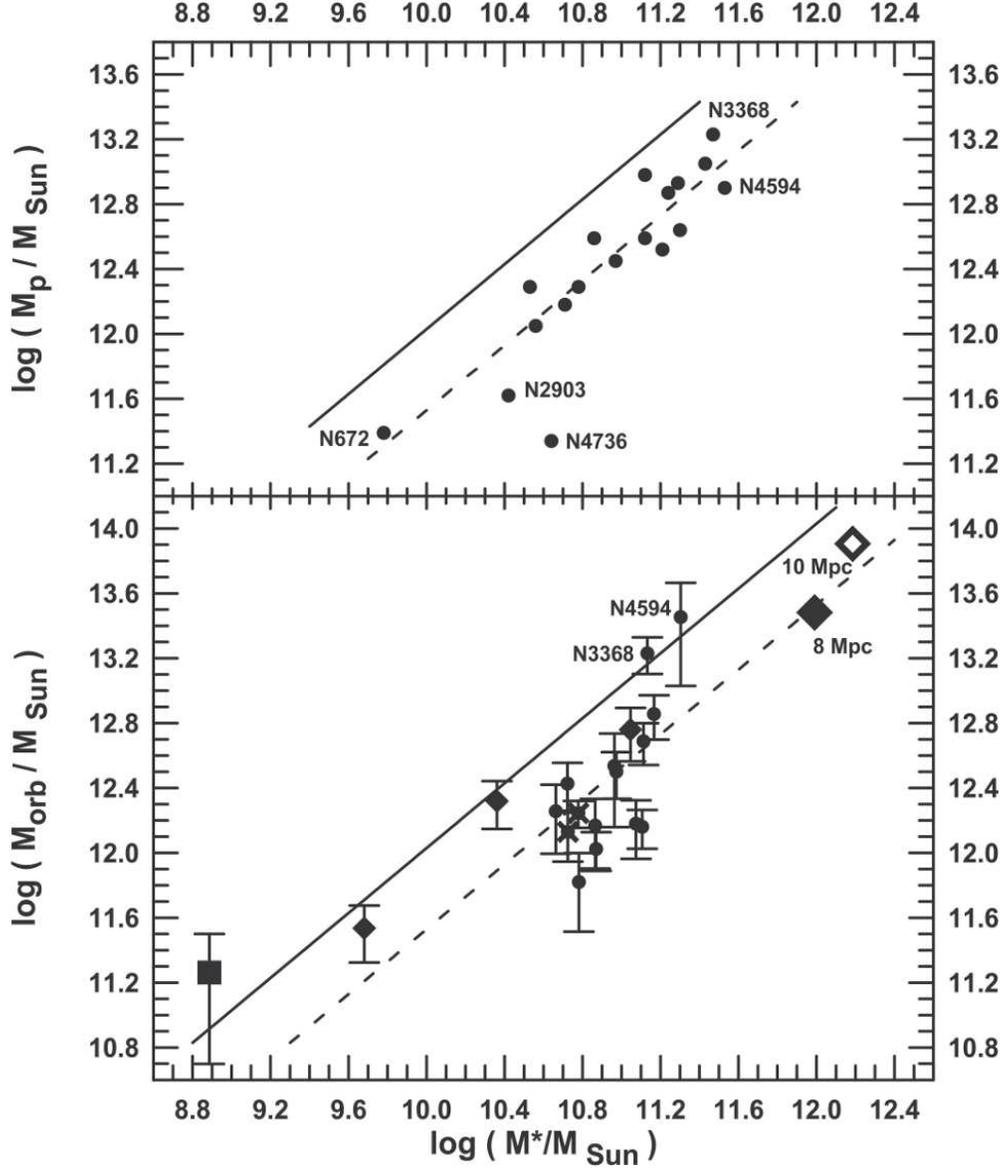}
\caption{Upper panel: the projected mass of nearby groups as a function
           of sum of their stellar mass. Some individual groups are marked
	   by the name of their principal member. The diagonal solid and dashed
           lines indicate the ratios: $M_{DM}/M_* = 97$ and 31, respectively.
	   Bottom panel: the orbital mass of nearby suites vs. sum of their
           stellar mass. Solid circles with error bars correspond to individual
	   rich suites. The  suites around the MW and M31 are marked by crosses.
	   Small solid diamonds with error bars indicate the weighted mean
	   ratios for the synthetic suites divided onto three sets: L (large),
           M (medium) and S (small) according to their Main Disturber mass.
	   The filled square indicates the weighted average ratio for 12 isolated
	   pairs of dwarfs.  The open and filled large diamonds show the total orbital
	   and stellar mass for the Local Volume of the radius of the 10 Mpc and 8 Mpc
	   radius, respectively. The diagonal lines mean the same as those in the upper panel.}
   \end{figure}

\end{document}